\documentstyle[aps,prl,eqsecnum,tighten,preprint,epsf]{revtex}
\begin{document}
\title{Zero temperature phase transitions in quantum Heisenberg
ferromagnets}
\author{Subir Sachdev and T. Senthil}
\address{Department of Physics, P.O.Box 208120,
	Yale University, New Haven, CT 06520-8120}
\date{February 5, 1996}
\maketitle

\begin{abstract}
The purpose of this work is to understand the zero temperature phases,
and the 
phase transitions, of Heisenberg spin systems which can have an
extensive,
spontaneous magnetic moment; this entails a study of quantum transitions
with an
order parameter which is also a non-abelian conserved charge. To this end,
we
introduce and study a new class of lattice models of quantum rotors. We
compute
their mean-field phase diagrams, and present continuum, quantum
field-theoretic descriptions  of their low energy properties in different
regimes. We argue that, in spatial dimension $d=1$, the phase transitions in
itinerant
Fermi systems are in the same universality class as the
corresponding transitions in certain rotor models. We discuss implications of our
results for itinerant fermions systems in higher $d$, and for other physical
systems.
\end{abstract}
\pacs{xxxxxxxxxx}
\widetext

\section{Introduction}
\label{intro}

Despite the great deal of attention lavished recently on magnetic quantum
critical phenomena, relatively little work has been done on systems in
which
one of the phases has an extensive, spatially averaged, magnetic moment.
In fact,  the
simple Stoner mean-field theory~\cite{stoner} of the zero temperature
transition from an
unpolarized Fermi liquid to a ferromagnetic phase
is an example of such a study, and is probably also
the earliest theory of a quantum phase
transition in any system.
What makes such phases, and the transitions between them, interesting is
that
the order parameter is also a conserved charge; in systems with a
Heisenberg $O(3)$
symmetry this is expected to lead to strong constraints on the critical
field
theories~\cite{conserve}.  As we will discuss briefly below, a number of
recent experiments
have studied systems in which
the quantum fluctuations of a ferromagnetic order parameter appear to
play
a central
role. This emphasizes the need for a more complete theoretical
understanding of quantum transitions  
into such phases.
In this paper, we will introduce what we believe are the
simplest theoretical models  which display phases and phase transitions
with these properties.
The degrees of freedom of these models are purely bosonic and consist of
quantum rotors on the
sites of a lattice. We will also present a fairly complete theory of the
universal properties of
the phases and phase transitions in these models, at least in spatial
dimensions
$d > 1$. Our quantum rotor models completely neglect charged and
fermionic
excitations and can
therefore probably be applied directly only to insulating ferromagnets.
However, in $d=1$, we
will argue that the  critical behaviors of transitions in metallic, fermionic
systems are  identical to
those of the corresponding transitions in certain quantum rotor models.

We now describe the theoretical and experimental motivation behind our
work:\\
({\em i}) The Stoner mean field theory of ferromagnetism~\cite{stoner} in    
         electronic systems in fact contains two transitions: one from an
unpolarized
Fermi liquid to a partially polarized itinerant ferromagnet (which has received some
recent experimental attention~\cite{lonzarich}), and the second from the
partially polarized to the saturated ferromagnet. A theory of fluctuations
near the
first critical point has been proposed~\cite{hertz,millis} but many basic
questions
remain unanswered~\cite{conserve}, especially on the ordered
side~\cite{scs} (there is
no proposed theory for the second transition, although we will outline one
in this
paper). It seems useful to examine some these issues in the simpler
context of
insulating ferromagnets. Indeed, as we have noted, we shall argue below
that in
$d=1$, certain insulating and itinerant systems have phase transitions
that are
in the same universality class.\\ 

({\em ii}) Many of the phases we expect
to
find in our model also exist in experimental compounds that realize the
so-called ``singlet-triplet'' model~\cite{st1}. These compounds were
studied
many years ago~\cite{st2} with a primary focus on finite temperature,
classical
phase transitions; we hope that our study will stimulate a re-examination
of
these systems to search for quantum phase transitions\\  ({\em iii}) All
of the
phases expected in our model (phases A-D in Section~\ref{sec:phases}
below),
occur in the
$La_{1-x} Sr_{x} Mn O_3$ compounds~\cite{lamno}. These, and related
compounds, have seen a great
deal of recent interest for their technologically important ``colossal
magnetoresistance''.\\ 
({\em iv}) Recent NMR experiments by Barrett {\em
et. al.}~\cite{barrett} have studied the magnetization of a quantum Hall
system  as a function of
both filling factor, $\nu$, and temperature, $T$, near $\nu =1$. The
$T=0$ state at $\nu = 1$ is a fully polarized 
ferromagnet~\cite{lk,fertig,sondhi,macdonald} and its finite temperature
properties have been studied from a field-theoretic point of
view~\cite{ferroqhe}.
More interesting for our purposes here is the physics away from $\nu=1$:
Brey {\em et. al.} have proposed a variational ground state consisting of a
crystal of
``skyrmions''; this state has magnetic order that is {\em canted}~\cite{nick},
{\em i.e.} in addition to a
ferromagnet moment, the system has magnetic order (with a vanishing
spatially averaged moment) in
the plane perpendicular to the average moment. A phase with just this
structure will appear in our
analysis, along with a quantum-critical point between a ferromagnetic
and a canted phase.
Although our microscopic models are quite different from those
appropriate for the quantum Hall
system, we expect the insights and possibly some universal features of
our results to be
applicable to the latter.

There has also been some interesting recent work on the effects of randomness
on itinerant ferromagnets~\cite{bk}. This paper shall focus exclusively on clean
ferromagnets, and the study of the effects of randomness on the models of this
paper remains an interesting open problem; we shall make a few remarks on this in
Section~\ref{randomness}

In the following subsection we will
introduce one of the models studied in this paper, followed by a brief description
of its phases in Section~\ref{sec:phases}. Section~\ref{intro} will conclude with an
outline of the remainder of the paper.

\subsection{The Model}
\label{sec:model}

We introduce the quantum rotor model which shall be the main focus of
the paper; extensions to related models will be considered later in the
body of the paper. On each
site $i$ of a regular lattice in
$d$ dimensions there is a
rotor whose configuration space is the surface of a sphere, described by
the
3-component unit vector $\hat{n}_{i\mu}$ ($\mu = 1,2,3$ and
$\sum_{\mu}
\hat{n}_{i\mu}^2 = 1$); the caret denotes that it is a quantum operator.
The canonically conjugate angular momenta are the
$\hat{L}_{i\mu}$, and these degrees of freedom obey the commutation
relations
(dropping the site index as all operators at different sites commute)
\begin{equation}
[\hat{n}_{\mu} , \hat{n}_{\nu}] = 0~~;~~
[\hat{L}_{\mu} , \hat{L}_{\nu}] = i\epsilon_{\mu\nu\lambda}
\hat{L}_{\lambda}~~;~~
[\hat{L}_{\mu} , \hat{n}_{\nu}] = i\epsilon_{\mu\nu\lambda}
\hat{n}_{\lambda}
\label{commute}
\end{equation}
As an operator on wavefunctions in the $n_{\mu}$ configuration space,
$\hat{L}_\mu$ is given by
\begin{equation}
\hat{L}_{\mu} = - i \epsilon_{\mu\nu\lambda} n_{\nu} \frac{\partial}{
\partial n_{\lambda}}
\label{lrep}
\end{equation}
We will be interested primarily in the properties of the Hamiltonian
\begin{equation}
\hat{H} = \frac{g}{2} \sum_i \left( \hat{L}_{i\mu}^2 + \alpha \left(
\hat{L}_{i\mu}^2 \right)^2 \right) - \sum_{<ij>} \left(
J \hat{n}_{i\mu} \hat{n}_{j\mu} + K \hat{L}_{i\mu} \hat{L}_{j\mu} + M
\left(
\hat{n}_{i\mu} \hat{L}_{j \mu} + \hat{n}_{j\mu} \hat{L}_{i\mu} \right)
\right)
\label{hrotor}
\end{equation}
where there is an implied summation over repeated $\mu$ indices,
and $<ij>$ is the sum over nearest neighbors, and the couplings $g$,
$\alpha$, $J$, $K$,
$M$ are all positive.
All previous analyses of quantum rotor
models~\cite{susskind,CHN,CSY,YSR} 
have  focussed exclusively on the case $K=M=0$, and the novelty
of our results arises
primarily from nonzero values of the new $K,M$ couplings. A crucial
property of $\hat{H}$ is that the
3 charges
\begin{equation}
\hat{Q}_{\mu} = \sum_i \hat{L}_{i\mu}
\end{equation}
commute with it, and are therefore conserved. Indeed, $\hat{H}$ is the
most
general Hamiltonian with
bilinear, nearest neighbor couplings between the $\hat{n}_{\mu}$ and
$\hat{L}_{\mu}$
operators, consistent
with conservation of the $\hat{Q}_{\mu}$. We have also included a single
quartic
term, with
coefficient $g
\alpha$, but its role is merely to suppress the contributions of
unimportant high energy
states.

Discrete symmetries of $\cal{H}$ will also be important
in our considerations. Time-reversal symmetry, ${\cal T}$ is realized
by the transformations
\begin{equation}
{\cal T}:~~~~~~\hat{L}_\mu \rightarrow - \hat{L}_\mu~~~~\hat{n}_\mu
\rightarrow - \hat{n}_\mu
\end{equation}
Notice that the commutators (\ref{commute}) change sign under ${\cal
T}$,
consistent with it being an anti-unitary transformation. 
All the models considered in this paper will have ${\cal T}$ as a
symmetry.
For the special
case $M=0$, we also have the additional inversion symmetry ${\cal P}$:
\begin{equation}
{\cal P}:~~~~~~\hat{L}_\mu \rightarrow \hat{L}_\mu~~~~\hat{n}_\mu
\rightarrow - \hat{n}_\mu
\label{cali}
\end{equation}
The presence of ${\cal P}$ will make the properties of the $M=0$
system somewhat different from the $M \neq 0$ case. We will see later
that
${\cal P}$ is related to a discrete spatial symmetry of the underlying
spin system that $\hat{H}$ models.

The utility of $\hat{H}$ does not lie in the possibility of finding an
experimental system
which may be explicitly modeled by it. Rather, we will find that it
provides a particularly
simple and appealing description of quantum phases and phase transitions
with a conserved order
parameter in a system with a non-abelian symmetry. Further, we will
focus
primarily on universal
properties of $\hat{H}$, which are dependent only on global symmetries of
the
states;  these
properties  are expected to be quite general and should apply also to other
models with the same
symmetries, including those containing ordinary Heisenberg spins.

To help the reader develop some intuition on the possibly unfamiliar
degrees of freedom in $\hat{H}$, we consider in Appendix~\ref{twolayer}
a
general
double-layer Heisenberg spin model~\cite{double} containing both inter-
and intra-layer
exchange interactions.
We show that, under suitable conditions, there is a fairly explicit mapping
of the double layer
model to the quantum rotor Hamiltonian $\hat{H}$. Under this mapping we
find that
each {\em pair\/} of
adjacent spins on the two layers behaves like a single quantum rotor.  In
particular $\hat{L}_{\mu}
\sim \hat{S}_{a\mu} + \hat{S}_{b\mu}$ and $\hat{n}_{\mu} \sim
\hat{S}_{a\mu}
 - \hat{S}_{b\mu}$ where
$a,b$ are the two layers
and the
$\hat{S}_{\mu}$ are Heisenberg spins. Notice also that ${\cal P}$ is
a layer-interchange symmetry.

Before turning to a description of the ground state of $\hat{H}$, it is
useful to
draw a parallel
to another model which has seen a great deal of recent interest---the
boson
Hubbard model~\cite{fwgf}.
The latter model has a single conserved charge, $\hat{N}_b$ the total
boson
number, associated with an
abelian global $U(1)$ symmetry. In contrast, the quantum rotor model
$\hat{H}$
has the 3 charges
$\hat{Q}_{\mu}$, and a non-abelian global $O(3)$ symmetry. As we will
see, the
non-abelian symmetry plays a
key role and is primarily responsible for the significant differences
between $\hat{H}$ and the boson
Hubbard model. 
It is also useful to discuss a term-by-term mapping between $\hat{H}$
and the
boson Hubbard model.
The terms
proportional to $g$ in $\hat{H}$ are analogous to the on-site  Hubbard
repulsion
in the boson model. The
latter model also has an on-site chemical potential term which couples
linearly to $N_B$, but such a
term is prohibited by symmetry in the non-abelian rotor model. The $J$
term
in $\hat{H}$ has an effect
similar to the boson hopping term, while the $K$ term is like a
nearest-neighbor boson
density-density interaction. There is no analog of the $M$ term in the
boson
Hubbard model.

\subsection{Zero temperature phases of $\hat{H}$}
\label{sec:phases}

\begin{figure}
\epsfxsize=5.5in
\centerline{\epsffile{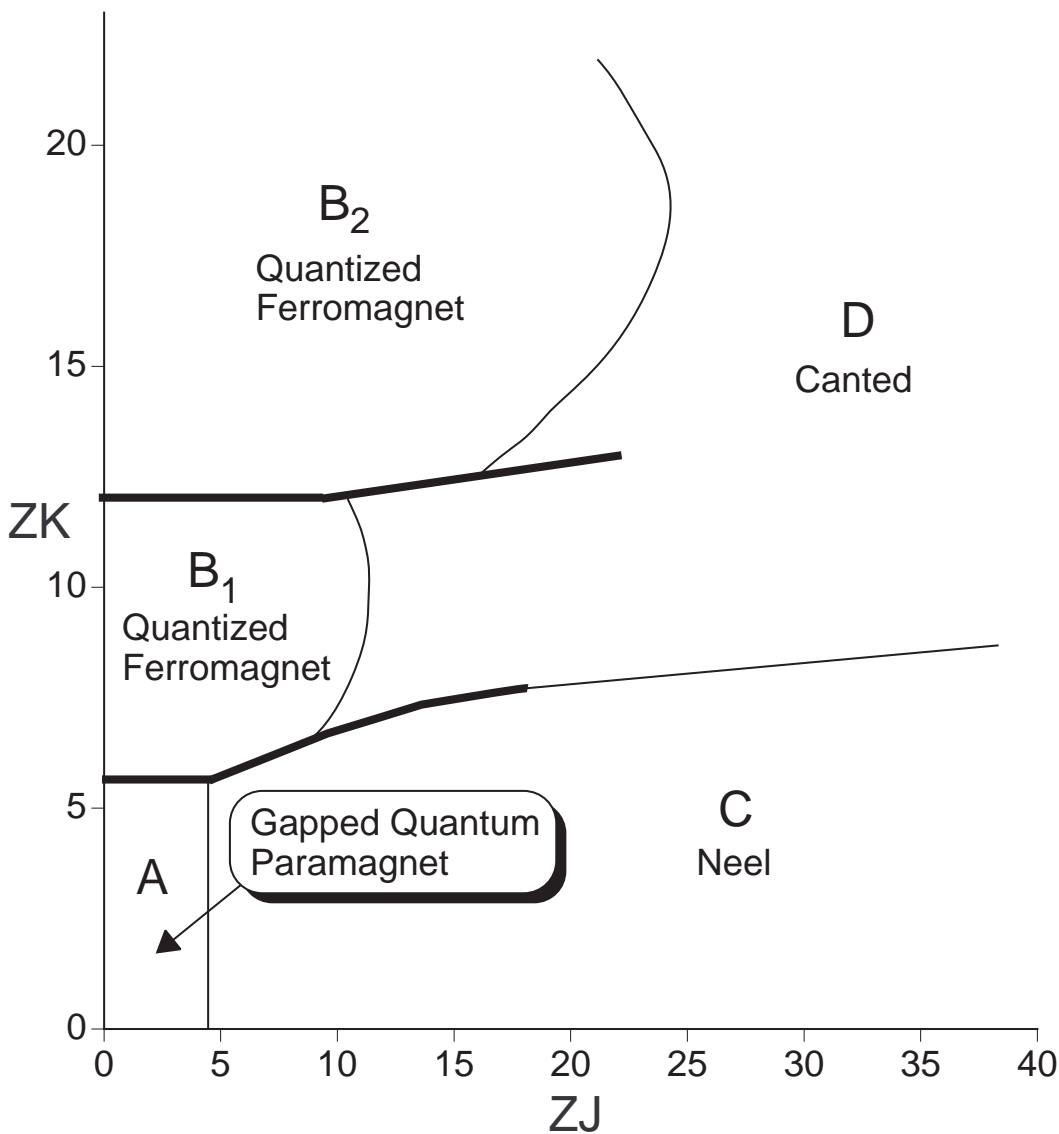}}
\vspace{0.3in}
\caption{Mean-field phase diagram of $\hat{H}$
(Eqn~\protect\ref{hrotor})
as a function of the
couplings
$J$ and $K$ at $M Z=4$, $g=1$, and $\alpha=1/2$; $Z$ is the co-ordination
number
of the lattice. 
The model does not have ${\cal P}$ symmetry at these values, but
the phase diagram for the ${\cal P}$ symmetric case is very similar. 
The mean-field
result becomes exact in the limit of large spatial dimensionality, but the general
features of the phase diagram are expected to hold for all $d>1$. Thin lines represent
second-order transitions while thick lines are first order. The quantized
ferromagnetic phases
$\rm{B}_\ell$ have magnetic moment per site
$\ell$;
there is an infinite sequence of these phases for all integers $\ell >0$
at larger values of $K$, and only the first two are shown.
The phases have the following ground state expectation values, up to a
global
$O(3)$ rotation:
(A) $\langle\hat{L}_\mu \rangle =0$, $\langle \hat{n}_\mu \rangle = 0$;
($\rm{B}_{\ell}$) $\langle\hat{L}_z \rangle = \ell $,
$\langle \hat{n}_z \rangle \neq 0$, $\langle\hat{L}_{x,y} \rangle =0$,
$\langle \hat{n}_{x,y} \rangle = 0$;
(C) $\langle\hat{L}_\mu \rangle = 0 $,
$\langle \hat{n}_z \rangle \neq 0$,
$\langle \hat{n}_{x,y} \rangle = 0$;
(D) $\langle\hat{L}_{x,z} \rangle \neq 0 $,
$\langle \hat{n}_{x,z} \rangle \neq 0$,
$\langle\hat{L}_{y} \rangle = 0 $,
$\langle \hat{n}_{y} \rangle = 0$.
}
\label{phasediag1}
\end{figure}

We show in Fig~\ref{phasediag1} the zero temperature ($T$) phase
diagram of
$\hat{H}$ in the $K,J$ plane
at fixed $g$, $\alpha$, and $M$.
This phase diagram was obtained using a mean-field theory which
becomes in
exact in the limit of
large spatial dimensionality ($d$); however, the topology and general
features are expected to be
valid for all $d > 1$. The $d=1$ case will be discussed separately later in
the paper; in the
following discussion we will assume $d>1$. We will also assume below that ${\cal
P}$ symmetry is absent, unless otherwise noted. Throughout this
paper we will  restrict consideration to parameters for which the ground state of
$\hat{H}$ are translationally invariant ground states---this will require that $M$
not be too large.

There are four distinct classes of phases:

\noindent
(A) \underline{Quantum Paramagnet:}

This is a featureless spin singlet and there is a gap to all excitations.
The $O(3)$ symmetry
remains unbroken, as 
\begin{equation}
\left\langle \hat{n}_{\mu} \right\rangle =
\left\langle \hat{L}_{\mu}
\right\rangle =0.
\end{equation}
Clearly this phase will always 
occur when $g$ is much bigger than all the other couplings.

\noindent
(B)\underline{Quantized Ferromagnets:}

These are ordinary ferromagnets in which the total moment of the ground
state is quantized in
integer multiples of the number of quantum rotors (extensions of
$\hat{H}$ in
which the quantization is
in half-integral multiples will be considered later in this paper).
The ferromagnetic order parameter
chooses a direction in spin space (say,
$z$), but the symmetry of rotations about this direction remains unbroken.
The ground state
therefore has the expectation values
\begin{equation}
\left\langle \hat{L}_z \right\rangle = \mbox{integer} \neq 0~~~~~;~~~~~
\left\langle \hat{n}_z \right\rangle \neq 0
\end{equation}
The value of $\left\langle \hat{n}_z \right\rangle$ is not quantized and
varies
continuously as $K$, $J$,
and $M$ are varied; for the system with ${\cal P}$ symmetry ($M=0$) we will have
$\left\langle \hat{n}_z \right\rangle = 0$ in this phase.
 If we
consider each quantum rotor as an effective degree of freedom
representing a set of  underlying Heisenberg spins (as in the double-layer
model of Appendix~\ref{twolayer}), then
$\left\langle \hat{n}_z \right\rangle$ determines the manner in which
the
quantized moment is
distributed among the constituent spins.
The low-lying excitation of these phases is a gapless, spin-wave
mode whose
frequency $\omega \sim k^2$ where $k$ is the wavevector of the
excitation.

That these phases occur is seen as follows: First consider the line
$J = 0, M = 0$. For very small $K$, it is clear that the ground state
is a quantum paramagnet.As $K$ is increased, it is easy to convince
oneself (by an explicit calculation) that a series of quantized
ferromagnet phases with increasing values of 
$\left\langle \hat{L}_z \right\rangle =$ integer get stabilized.
Further in this simple limit the exact ground state in each one 
of these phases is just a state in which each site is put
in the same eigenstate of $L^{2}$ and $L_z$. There is a finite energy
cost to change the value of $L^{2}$ at any site. Now consider
moving away from this limit by introducing 
small non-zero values of 
$J$ and $M$. These terms vanish in the subspace of states
with a constant value of $L^{2}_{i}$ so we need to consider
excitations to states which involve changing the value of
$L^{2}_{i}$ at some site. As mentioned above such states are
separated from the ground state by a gap. Consequently, though
the new ground state is no longer the same as at $J = 0, M = 0$,
it's quantum numbers , in particular the value of 
$\hat{Q}_z  = \sum_i \hat L_{iz} $, are unchanged.The stability of the 
quantized ferromagnet phases up to finite values of $J$ and $M$
implies the existence of direct transitions between them
which is naturally first order.
In general a non-zero value of 
$M$ will also lead to a non-zero value of $\left\langle \hat n_z
\right\rangle$.

All of this  should remind
the reader of the
 Mott-insulating phases of the boson Hubbard model~\cite{fwgf}. In the
latter, the boson number,
$\hat{n}_b$, is quantized in integers, which is the analog of the conserved
angular momentum $\hat{L}_{\mu}$ of
the present model. However, the properties of the Mott phases are quite
similar for all values
of $\left\langle \hat{n}_b \right\rangle $, including $\left \langle
\hat{n}_b \right\rangle =0$. In contrast, for the $O(3)$ rotor model, the
case $\left\langle
\hat{L}_z \right\rangle =0$ (the quantum paramagnet,
which has no broken symmetry and a gap to all
excitations)
is quite different from
$\left\langle \hat{L}_z \right\rangle = \mbox{integer} \neq 0$ cases (the
quantized
ferromagnets, which have a broken symmetry and
associated gapless spin-wave excitations). As we argued above and
as shown in Fig~\ref{phasediag1}, the lobes of the quantized
ferromagnets and the quantum
paramagnet are separated by first-order transitions, while there were no
such transitions in the
phase diagram for the boson Hubbard model in Ref~\cite{fwgf}; this is an
artifact of the absence
of off-site boson attraction terms analogous to 
the $K$ term in Ref~\cite{fwgf}, and is {\em not\/} an
intrinsic difference
between the abelian and non-abelian cases.

\noindent
(C) \underline{ N\'{e}el Ordered Phase:}

In this phase we have 
\begin{equation}
\left\langle \hat{n}_{\mu} \right\rangle \neq 0, \mbox{ while
}\left\langle
\hat{L}_{\mu} \right\rangle = 0. 
\end{equation}
This occurs when $J$ is 
much bigger than all other couplings. We refer to it as the
N\'{e}el phase because the spins in the bilayer model are oriented in
opposite directions in the two
layers. More generally, this represents any phase in which the
spin-ordering is defined by a single
vector field ($n_{\mu}$) and which has no net ferromagnetic moment.
There
are  2 low-lying
spin-wave modes, but they now have a linear dispersion
$\omega \sim k$~\cite{tops}.
The order parameter condensate $\left\langle n_{\mu}
\right\rangle$ does not have a
quantized value and varies continuously with changes in the couplings.

From the perspective of classical statistical mechanics, the existence of
this
N\'{e}el phase is rather
surprising. Note that there is a linear coupling $M \hat{L}_{i\mu}
\hat{n}_{i\mu}$ in
$\hat{H}$ between the
$\hat{n}_{\mu}$ and $\hat{L}_{\mu}$ fields. In a classical system, such a
coupling
would imply
that a non-zero condensate of $\hat{L}_{\mu}$ must necessarily
accompany any
condensate in $\hat{n}_{\mu}$.
The presence of a phase here, in which $\left\langle \hat{L}_{\mu}
\right\rangle =
0$ despite
$\left\langle \hat{n}_{\mu} \right\rangle \neq 0$, is a consequence of the
quantum
mechanics of the
conserved field $\hat{L}_{\mu}$.

\noindent
(D) \underline{Canted Phase:}

Now both fields have a condensate:
\begin{equation}
\left\langle \hat{n}_{\mu} \right\rangle \neq 0 \mbox{ and }\left\langle
\hat{L}_{\mu} \right\rangle \neq 0. 
\end{equation}
The magnitudes of, and
relative
angle between, the condensates can all take arbitrarily values, which vary
continuously as the
couplings change; however if ${\cal P}$ is a symmetry (if $M=0$)
then $\left\langle \hat{n}_{\mu} \right\rangle$ is always orthogonal to
$\left\langle \hat{L}_{\mu} \right\rangle$.
 The $O(3)$
symmetry of
$\hat{H}$ is completely broken as it takes two vectors to
specify the orientation of the condensate. In the bilayer model, the spins
in the two layers
are oriented in two non-collinear directions, such that there is a net
ferromagnetic moment.
The low lying excitations of this phase consist of 2 spin wave modes, one
with $\omega \sim k$,
while the other has $\omega \sim k^2$.
This phase is the analog of the superfluid phase in the
boson Hubbard model. In the latter model the conserved number density has
an arbitrary,
continuously varying value and there is long range order in the conjugate
phase variable;
similarly here the conserved
$\left\langle \hat{L}_{\mu} \right\rangle$ has an arbitrary value and the
conjugate $\hat{n}_{\mu}$ field
has a definite orientation.

At this point, it is useful to observe a parallel between the phases of
the quantum rotor model and those of a ferromagnet Fermi liquid. This
parallel is rather crude
for $d>1$, but, as we will see in Secs~\ref{deq1} and~\ref{conc}, it can be
made
fairly explicit
in $d=1$ where we
believe that the universality classes of the transitions in the rotor model
and the Fermi liquid
are identical.
The quantized ferromagnetic phases B are the analogs of the fully
polarized
phase of the Fermi
liquid in which the spins of all electrons are parallel. The canted phase D
has a continuously
varying ground state polarization, as in a partially polarized Fermi
liquid. Finally the
N\'{e}el phase C is similar to an unpolarized Fermi liquid in that both
have no net
magnetization and exhibit gapless spin excitations.

Finally, we point out an interesting relationship between phases with a
non-quantized value of
the average magnetization (as in phase D of the rotor model), and phases
with
vanishing average magnetization (phases A and C). Notice that, in both the
Fermi liquid and the quantum rotor model, it is possible to have a
continuous
transition between such phases.
However, in both cases, such a transition only occurs when the phase
with
vanishing magnetization has gapless spin excitations; in phase C of the
rotor model we have the gapless spin waves associated with the N\'{e}el
long
range order, while in the unpolarized Fermi liquid we have gapless 
spin carrying fermionic quasiparticles.
Phase A of the rotor model has no gapless excitations, but notice that
there is no continuous
transition between it and the partially polarized canted phase D.
We believe this feature to be a general principle: {\em continuous zero
temperature transitions in
which there is an onset in the mean value of a non-abelian conserved
charge
only occur from phases
which have gapless excitations.}

The outline of the remainder of the paper is as follows. We will begin in
Section~\ref{mft} by discussing the mean field theory which produced the 
phases described above. In Section~\ref{struct} we will perform a small
fluctuation analysis of the low-lying excitations of the phases.
We will turn our attention to the quantum phase transitions in $d>1$ in
Section~\ref{qpt}, focusing mainly on the transition between phases C and D in
Section~\ref{CtoD}, and that between phases B and D in Section~\ref{BtoD}.
In Section~\ref{monopole} we shall consider an extension of the basic rotor model
(\ref{hrotor}): each rotor will have a `magnetic monopole' at the origin of $n$
space, which causes the angular momentum of each rotor to always be non-zero.
We will turn our attention to the important and distinct physics in $d=1$ in
Section~\ref{deq1}. We will conclude in Section~\ref{conc}  by placing our results
in the context of earlier work and discuss future directions for research. Some
ancillary results are in 5 appendices: we note especially Appendix~\ref{appcalc},
which contains new universal scaling functions of the  dilute
Bose gas, a model which turns out to play a central role in our analysis.

\section{Mean field theory}
\label{mft}

In this section we will describe the mean-field theory in which the
phase diagram of Fig~\ref{phasediag1} was obtained. The mean-field
results
become exact in the limit of large spatial dimensionality, $d$. Rather
than explicitly discussing the structure of the large $d$ limit, we choose
instead a more physical discussion. We postulate on every site
a single-site mean field
Hamiltonian
\begin{equation}
\hat{H}_{mf} = \frac{g}{2} \left( \hat{L}_{\mu}^2 + \alpha \left(
\hat{L}_{\mu}^2 \right)^2 \right) - N_{\mu} \hat{n}_{\mu}
- h_{\mu} \hat{L}_{\mu}
\label{hmf}
\end{equation}
which is a function of the variational c-number local fields
$N_{\mu}$ and $h_{\mu}$. These fields are determined at $T=0$
by minimizing the expectation value of $\hat{H}$ in the ground state
wavefunction of $H_{mf}$. The mean-field ground state energy of
$\hat{H}$ is
\begin{equation}
E_{mf} = E_0  - \frac{J Z}{2} \left\langle \hat{n}_{\mu} \right\rangle_0^2
- \frac{K Z}{2} \left\langle \hat{L}_{\mu} \right\rangle_0^2
- M Z \left\langle \hat{n}_{\mu} \right\rangle_0
\left\langle \hat{L}_{\mu} \right\rangle_0 + N_{\mu} \left\langle
\hat{n}_{\mu} \right\rangle_0 + h_{\mu} \left\langle
\hat{L}_{\mu} \right\rangle_0
\end{equation}
where $E_0$ is the ground state energy of the $\hat{H}_{mf}$, all
expectation
values are in the ground state wavefunction of $\hat{H}_{mf}$, and $Z$ is
the
co-ordination number of the lattice. We now have to minimize
the value of $E_{mf}$ over variations in $h_{\mu}$ and $N_{\mu}$.
This was carried out numerically for a characteristic set of values of the
coupling
constants. Stability required that the coupling $M$ not be too large.
Further details may be
found in Appendix~\ref{mfta}. Here we describe the behavior of the local
effective fields $h_{\mu}$, $N_{\mu}$ in the various phases:\\
(A) {\em Quantum Paramagnet:}\\
This phase has no net effective fields $N_{\mu} = 0$, $h_{\mu}  = 0 $.\\
(B) {\em Quantized Ferromagnets:} \\
We now have $N_z \neq 0$ and $h_z \neq 0$ with all other components
zero.
The values of $N_z$ and $h_z$ both vary continuously as the parameters
are
changed. Nevertheless, the  value of $\left\langle \hat{L}_z \right
\rangle$
remains pinned at a fixed non-zero integer. This is clearly possible only
because $\hat{L}_{\mu}$ commutes with $\hat{H}$ and $\hat{H}_{mf}$,
and $L_z$ is
therefore a good quantum number. \\
(C) {\em N\'{e}el Ordered Phase:} \\
Like phase B, this phase has $N_z \neq 0$ and $h_z \neq 0$ with
all other components zero, and the
values of $N_z$ and $h_z$ both vary continuously as the parameters are
changed. However $\left\langle \hat{L}_z \right \rangle = 0 $; it
is now quantized at
an integer value which happens to be zero. As a result, there is no net
ferromagnetic moment in this phase. This unusual relationship between
an order parameter $\left\langle \hat{L}_{\mu} \right\rangle$, and
its conjugate field $h_{\mu}$, is clearly a special property of the
interplay
between quantum mechanics and conservation laws, and cannot exist in
classical statistical mechanics systems.\\
(D) {\em Canted Phase:}\\
Now both $h_{\mu}$ and $N_{\mu}$ are non-zero, and take smoothly
varying
values with no special constraints, as do their conjugate fields
$\hat{n}_{\mu}$ and $\hat{L}_{\mu}$. In systems with ${\cal P}$ a good
symmetry
($M=0$), we have $\sum_{\mu} h_{\mu} N_{\mu} = 0$.

\section{Structure of the Phases}
\label{struct}

The quantum paramagnet A is a featureless singlet phase with all
correlations
decaying exponentially in both space and imaginary time, and a gap to
all excitations. The ferromagnet phases B and the N\'{e}el phase C
are conventional magnetically ordered phases and hardly need further
comment here. We describe below the long-wavelength, low energy
quantum
hydrodynamics of the canted phase D.
We are implicitly assuming here, and in the remainder of this section that
$d>1$.

We will study the phase D by accessing it from the N\'{e}el phase C. We
will
analyze the properties of D for small values of the uniform ferromagnetic
moment;
this leads to a considerable simplification in the analysis, but the form of
the results are quite general and hold over the entire phase D---in a later
section (Section~\ref{BtoD}), we will also access phase D from one of the
quantized
ferromagnetic phases B and obtain similar results.

We will use an imaginary time,
Lagrangian based functional-integral point of view.
The analysis begins by decoupling the inter-site interactions in $\cal{H}$
by
the spacetime dependent
Hubbard Stratonovich fields $N_{\mu} (x, \tau) $ and $h_{\mu} (x, \tau)$;
these
fields act as dynamic local fields similar to those in $\hat{H}_{mf}$
(Eqn~(\ref{hmf})). We can then set up the usual Trotter product
decomposition
of the quantum mechanics independently on each site: this yields the
following
local functional integral on each site (we are not displaying the inter-site
terms involving the $N_{\mu}$ and $h_{\mu}$ fields, as these will be
considered later):
\begin{displaymath}
Z_L = \int {\cal D} n_{\mu} \delta ( \vec{n}_{\mu}^2 -1 ) \exp\left( -
\int_0^{\beta} d
\tau {\cal L}_L \right)
\end{displaymath}
\begin{equation}
{\cal L}_L = \frac{1}{2g} \left( \frac{\partial n_{\mu}}{\partial \tau}
- i \epsilon_{\mu\nu\lambda} h_{\nu} n_{\lambda} \right)^2 - n_{\mu}
N_{\mu}
\label{zloc}
\end{equation}
We have ignored, for simplicity, the contribution of the quartic $\alpha$
term
in $\cal{H}$. It is not possible to evaluate $Z_L$ exactly; for
time-independent source fields $h_{\mu}$, $N_{\mu}$, the evaluation of
$Z_L$ is
of course equivalent to the numerical diagonalization that was carried out
in Section~\ref{mft}. For $N_{\mu} = 0$, however, we can obtain the
following
simple formula for the ground state energy $E_L = - \lim_{\beta
\rightarrow
\infty} (1/\beta) \log Z_L$:
\begin{equation}
E_L = \mbox{Min} \left[ g \ell (\ell + 1) /2 - \ell h \right]~~~~; \ell \geq
0
~,~\mbox{integer}.
\label{eloc}
\end{equation}
The minimum is taken over the allowed values of $\ell$, and we have again
ignored the $\alpha$ term. Note that this is a highly non-analytic function
of
$h$: these non-analyticities are directly responsible for the
lobes of the quantized ferromagnetic phases B. In this section we will
begin
by working in the region of parameters in which $E_L$ is minimized
by $\ell = 0$ (phases A and C);
in the vicinity of this region it is permissible to expand in powers of $h$.
Inter-site effects will then eventually lead to a phase in which there is a
net
uniform moment (phase D)--- this moment will not be quantized and there
will be
appreciable fluctuations in the magnetic moment of each site (this is
similar to
fluctuations in particle number in a boson superfluid phase).
For simplicity we will first present the analysis for the case
$M = 0$. Later we will indicate the modifications necessary when
$M \neq 0$. 

\subsection{Model with ${\cal P}$ symmetry}
\label{secmeq0}
Recall that ${\cal P}$ symmetry is present when $M=0$.

It is convenient to write an effective action functional in terms
of the $h_\mu$ and $n_\mu$ fields:
 the form of this functional can be guessed by symmetry
and the usual Landau arguments:
\begin{displaymath}
Z = \int {\cal D} n_{\mu} {\cal D} h_{\mu}
\exp\left( - \int d^d x
\int_0^{\beta} d
\tau ({\cal L}_1 + {\cal L}_2 )\right)
\end{displaymath}
\begin{eqnarray}
{\cal L}_1 = && \frac{K_1}{2} \left( \frac{\partial n_{\mu}}{\partial \tau}
- i \epsilon_{\mu\nu\lambda} ( h_{\nu} + H_{\nu}) n_{\lambda} \right)^2
+ \frac{K_2}{2} (\nabla n_{\mu})^2
+ \frac{K_3}{2} (\nabla h_{\mu})^2 \nonumber \\
&&~~~~~~~~~~~~~~~~~~~~~~
+ \frac{r_1}{2} h_{\mu}^2  + \frac{u_1}{8} (h_{\mu}^2)^2
\nonumber
\end{eqnarray}
\begin{equation}
{\cal L}_2 =  \frac{r_2}{2} n_{\mu}^2  + 
 + \frac{u_2}{8} (n_{\mu}^2)^2
+ \frac{v_1}{2} (n_{\mu}^2)(h_{\nu}^2) + \frac{v_2}{2} (n_{\mu} h_{\mu})^2
\label{landau}
\end{equation}
We have temporarily modified $n_{\mu}$ from a fixed-length to a ``soft-
spin'' field;
this is merely for convenience in the following discussion and not
essential.
Here $H_{\mu}$ is an external magnetic field whose coupling to the
fields is determined by gauge-invariance~\cite{conserve}. 
The reason for
splitting the Lagrangian into pieces ${\cal L}_1$, ${\cal L}_2$ will
become
clear below.

We can now look for static, spatially uniform, saddle points of ${\cal L}$.
This gives three different types of solutions, corresponding to the phases
A, C, and D (the absence of a length constraint on $n_{\mu}$ is necessary
to obtain
all three saddle-points at tree level). The
values of the
$h_{\mu}$ and
$n_{\mu}$ fields at these saddle points are identical in form to those
discussed for
these phases in Section~\ref{mft} (with the reminder that we have
temporarily
specialized to $M = 0$ so that ${\cal P}$ is a good symmetry). Note that
there is no
saddle point corresponding to the B phases: this is clearly a consequence
of ignoring
the non-analytic behavior of $h$ in $Z_L$.

The low-lying excitation spectrum in the A, C, and D phases can now be
determined by an analysis of Gaussian fluctuations about the saddle
points.
While simple in principle, such an analysis is quite tedious and involved,
especially in phase D. We will therefore not present it here; we present
instead a more elegant approach in which the answer can be obtained with
minimal
effort.

Recall that an efficient method of obtaining the properties of the N\'{e}el
phase is to use a non-linear sigma model in which the constraint
$n_{\mu}^2 = 1$ is imposed. This eliminates high energy states from the
Hilbert
space associated with amplitude fluctuations, but does not modify the low
energy spectrum. We introduce here an extension to a hybrid sigma model
which
allows also for the existence of the canted phase D. Notice from the mean-
field
solutions in Sec~\ref{mft} that the onset of the canted phase D is signaled
by the appearance of an expectation value of $h_{\mu}$ in a direction
orthogonal to the mean direction of $n_{\mu}$, while the component of
$h_{\mu}$
parallel to $n_{\mu}$ is zero in both the C (N\'{e}el) and D phases
(note: the last restriction on components of $h_{\mu}$ and $n_{\mu}$
parallel to
each other requires ${\cal P}$ symmetry and $M = 0$). This suggests that
the important
fluctuations are the components of
$h_{\mu}$ perpendicular to $n_{\mu}$, while fluctuations which change
the dot
product $n_{\mu} h_{\mu}$ are high energy modes. So we define our hybrid
sigma model by imposing the two rotationally invariant constraints
\begin{equation}
n_{\mu} n_{\mu} = 1~~~~~~;~~~~~~n_{\mu} h_{\mu} = 0
\label{constraints}
\end{equation}
With these constraints, the degrees of freedom have been reduced from
the
original 6 real fields $h_{\mu}$, $n_{\mu}$ to 4. Notice that while
amplitude
fluctuations of $n_{\mu}$ have been eliminated, those in the components
of $h_{\mu}$ orthogonal to $n_{\mu}$ have not---this is the reason for
the
nomenclature `hybrid' above. As an immediate consequence
of (\ref{constraints}), all the terms in ${\cal L}_2$ either become
constants
or modify couplings in ${\cal L}_1$,
and ${\cal L}_1$ is the Lagrangian of the hybrid sigma model.

The Lagrangian ${\cal L}_1$ and the constraints (\ref{constraints})
display, in principle, all three phases A, C, and D. We now move well away
from phase A, assuming that $n_{\mu}$ has a well developed expectation
value
along the $z$ direction. We then parametrize deviations from this state
by the following parametrization of the fields, which explicitly obeys the
constraints (\ref{constraints}):
\begin{eqnarray}
\vec{n} &=& \left( \frac{\psi + \psi^\ast}{\sqrt{2}}, \frac{\psi -
\psi^\ast}{\sqrt{2}i}, (1 - 2 |\psi|^2)^{1/2} \right) \nonumber \\
\vec{h} &=&  
\left( \frac{\phi + \phi^\ast}{\sqrt{2}}, \frac{\phi -
\phi^\ast}{\sqrt{2}i},
-\frac{\psi^\ast \phi + \phi^\ast \psi}{ ( 1 - 2 |\psi|^2 )^{1/2}} \right).
\label{paramet}
\end{eqnarray}
We have two complex fields $\psi, \phi$, corresponding to the 4 degrees
of
freedom in the hybrid sigma model. The field $\psi$ represents
oscillations
of the $n_{\mu}$ about its mean value: by definition,
we will have $\langle \psi \rangle = 0$ in all phases. The field $\phi$
measures
the amplitude of $h_{\mu}$ orthogonal to the instantaneous value of
$n_{\mu}$: we
have
$\langle \phi \rangle =0$ in the N\'{e}el phase C, while
$\langle \phi \rangle \neq 0$ in the canted phase D. We now insert
(\ref{paramet})
into (\ref{landau}) and obtain at $H=0$ the partition function $Z_{\sigma}$
for our hybrid
sigma model
\begin{eqnarray}
&&~~~~~~~~~Z_{\sigma} = \int {\cal D} \phi {\cal D} \phi^\ast
\frac{ {\cal D}
\psi  {\cal D} \psi\ast}{1 - 2 |\psi|^2} \exp \left( - \int d^d x
\int_0^{\beta}
d \tau ( {\cal L}_{\sigma 1} + {\cal L}_{\sigma 2} + \cdots ) \right)
\nonumber \\
&&~~~~{\cal L}_{\sigma 1} = K_1 \left| \frac{\partial \psi}{\partial \tau}
\right|^2
+ K_1 \left( \phi^{\ast} \frac{\partial \psi}{\partial \tau} -  \phi
\frac{\partial \psi^{\ast}}{\partial \tau} \right)
+ K_2 |\nabla \psi|^2 + K_3 |\nabla \phi|^2 + 
 r_4 |\phi|^2
\nonumber \\
&&~~~~ {\cal L}_{\sigma 2} = \frac{u_1}{2} 
|\phi|^4 + \frac{K_1}{2} \left( \frac{\partial
|\psi|^2}{\partial
\tau}
\right)^2 + K_1 |\psi|^2 \left( \phi^{\ast} \frac{\partial \psi}{\partial
\tau} -  \phi\frac{\partial \psi^{\ast}}{\partial \tau} \right)
+ \frac{K_2}{2} \left( \nabla |\psi|^2 \right)^2 \nonumber \\
&&~~~~~~~~~~~~~~~~~~~+ \frac{K_3}{2}
\left| \nabla( \psi^\ast \phi + \psi \phi^\ast ) \right|^2  
+ \frac{r_4}{2} (\psi^\ast \phi +
\psi \phi^{\ast} )^2
\label{lsigma1}
\end{eqnarray}
where $r_4 = r_1 + v_1 - K_1 $.
The Lagrangian ${\cal L}_{\sigma 1}$ (${\cal L}_{\sigma 2}$) contains
terms
that are quadratic (quartic) in the fields. Note that, apart from the $u_1$
term, all coupling constants in ${\cal L}_{\sigma 2}$ are related to those
in
${\cal L}_{\sigma 1}$: these constraints on the couplings encapsulate
the rotational invariance of the underlying physics. The field $\psi$,
which
represents fluctuations of $N_{\mu}$ about its average value has no
``mass'' term (a term with no gradients) in ${\cal L}_{\sigma 1}$; the
constraints
on the couplings ensure that no such mass term is ever generated, and that
the
fluctuations of $\psi$ remain gapless. This is exactly what is expected
as the N\'{e}el order parameter is non-zero in both the C and D phases.
Finally, note that although we have only displayed $Z_{\sigma}$ for $H=0$,
the form of the coupling to a field $H_z$ in the $z$ direction can be easily
deduced from the gauge invariance arguments of Ref~\cite{conserve}.

The transition between
phases C and D is controlled at tree-level by the sign of $r_4$, and phase
C is
present for $r_4 > 0$. In this case, we can integrate out the massive field
$\phi$, order by order, and obtain an effective action for $\psi$.
At tree-level, this effective action corresponds to putting $\phi=0$
in ${\cal L}_{\sigma}$. The $\psi$ fluctuations then represent the two
real
gapless spin wave modes of the N\'{e}el phase with dispersion
$\omega = \sqrt{K_2/K_1} k$. The loop corrections from the $\phi$
fluctuations
will simply renormalize the values of $K_2$ and $K_1$.

Finally, we turn to phase D which exists for $r_4 < 0$.
In this case, $\phi$ acquires a mean value
\begin{equation}
\langle \phi \rangle = \sqrt{\frac{-r_4}{ u_1 }}.
\end{equation}
We have chosen $\langle\phi\rangle$ to be real, corresponding to choosing
a
definite direction for the components of $h_{\mu}$ orthogonal to
$n_{\mu}$.
We now look at small fluctuations about this state by writing
\begin{eqnarray}
\phi &=& \sqrt{\frac{-r_4}{ u_1 }} + (\phi_x + i \phi_y)/\sqrt{2}
\nonumber\\
\psi &=& (\psi_x + i \psi_y)/\sqrt{2}
\end{eqnarray}
where all fields on the right hand side are real. The field $\phi_x$
represents changes in the magnitude of the condensate; its fluctuations
are
therefore massive and can be neglected. We Fourier transform the
remaining
fields,
introduce the vector $\Phi = (\psi_x, \psi_y, \phi_y)$ and obtain from
${\cal L}_1$ the quadratic action:
\begin{equation}
\int \frac{d^d k d\omega}{(2 \pi)^{d+1}} \frac{1}{2} \Phi_{i} (-k, -\omega)
M_{ij} (k, \omega ) \Phi_j (k, \omega )
\end{equation}
where the dynamical matrix $M$ is
\begin{equation}
M (k, \omega ) = \left(
\begin{array}{ccc}
K_1 \omega^2 + K_2 k^2 & 0 & K_1 \omega \\
0 & K_1 \omega^2 + K_2 k^2 & 0 \\
-K_1 \omega & 0 & K_3 k^2
\end{array} \right)
\end{equation}
The zeros of the determinant of $M$ give us the eigenfrequencies of the
harmonic
oscillations about the ground state in phase D. Analytically continuing to
real frequencies, we obtain one linearly dispersing mode
$\omega = \sqrt{K_2/K_1}k$ and one quadratically dispersing mode (at
small $k$)
$\omega = \sqrt{(K_3 K_2)/(K_1^2 + K_1 K_3 k^2)} k^2$.
Although there were three real fields to begin with, there are only two
distinct modes,
as a pair of them behave like canonically conjugate degrees of freedom.

\subsection{Model without ${\cal P}$ symmetry}
\label{secmneq0}
Now $M\neq 0$, and a new class of terms will be
permitted in
the effective action.

The analysis in this case is technically similar to the $M = 0$
case; so we will be very brief. The effective action functional for the
$n_\mu$ and
$h_\mu$ now becomes:
\begin{displaymath}
Z = \int {\cal D} n_{\mu} {\cal D} h_{\mu} 
\exp\left( - \int d^d x
\int_0^{\beta} d
\tau ({\cal L}_1 + {\cal L}_2 )\right)
\end{displaymath}
\begin{eqnarray}
{\cal L}_1 = && \frac{K_1}{2} \left( \frac{\partial n_{\mu}}{\partial \tau}
- i \epsilon_{\mu\nu\lambda} ( h_{\nu} + H_{\nu}) n_{\lambda} \right)^2
+ \frac{K_2}{2} (\nabla n_{\mu})^2
+ \frac{K_3}{2} (\nabla h_{\mu})^2 \nonumber \\
&&~~~~~~~~~~~~~~~~~~~~~~+ K_4 (\nabla n_{\mu}) (\nabla h_{\mu})
+ \frac{r_1}{2} h_{\mu}^2  + \frac{u_1}{8} (h_{\mu}^2)^2
\nonumber
\end{eqnarray}
\begin{equation}
{\cal L}_2 =  \frac{r_2}{2} n_{\mu}^2  + r_3 h_{\mu}
n_{\mu} + \frac{u_2}{8} (n_{\mu}^2)^2
+ \frac{v_1}{2} (n_{\mu}^2)(h_{\nu}^2) + \frac{v_2}{2} (n_{\mu} h_{\mu})^2
\label{landauM}
\end{equation}
Note the presence of the additional couplings $K_4$ and $r_3$; these
couplings were
forbidden earlier by the ${\cal P}$ symmetry.

We can again look for static, spatially uniform, saddle points of ${\cal L}$.
As before,
this gives three different types of solutions, corresponding to the phases
A, C, and D but none corresponding to the B phases. The low-lying
excitation spectrum can again be found by defining a hybrid sigma model
by imposing the constraints
\begin{equation}
n_\mu n_\mu = 1~~~~~~;~~~~~~n_\mu h_\mu = c
\label{constraintsM}
\end{equation}
where $c$ is some constant. Clearly the only difference from the previous
section is the non-zero (but constant) value of $n_\mu h_\mu$.
This is necessary  because though the onset of the canted phase D is still 
signaled by the appearance of an expectation value of $h_{\mu}$ in a
direction
orthogonal to the mean direction of $n_{\mu}$, the component of
$h_{\mu}$
parallel to $n_{\mu}$ is non-zero in both the C (N\'{e}el) and D phases
when $M \neq 0$ .

 We proceed as before and move well away
from phase A, assume that $n_{\mu}$ has a well developed expectation
value
along the $z$ direction, and replace the earlier parametrization
(\ref{paramet})
by the following: 
\begin{eqnarray}
\vec{n} &=& \left( \frac{\psi + \psi^\ast}{\sqrt{2}}, \frac{\psi -
\psi^\ast}{\sqrt{2}i}, (1 - 2 |\psi|^2)^{1/2} \right) \nonumber \\
\vec{h} &=&  c \vec{n} +
\left( \frac{\phi + \phi^\ast}{\sqrt{2}}, \frac{\phi -
\phi^\ast}{\sqrt{2}i},
-\frac{\psi^\ast \phi + \phi^\ast \psi}{ ( 1 - 2 |\psi|^2 )^{1/2}} \right).
\label{parametM}
\end{eqnarray}
The new feature is the first term in the second equation. From now on
the analysis is similar to that in the previous section, so we merely
state the results. At the Gaussian level, there now is a new term in the 
lagrangian ${\cal L}_{\sigma1}$ equal to $K_5(\nabla \psi^\ast\nabla \phi
+ \nabla \psi \nabla \phi^\ast)$. Both the N\'{e}el and canted phases 
continue to have the same excitation spectrum as at $M = 0$, but the 
constants of proportionality in the low $k$ dispersion relations
are different from their $M = 0$ values, as are the actual eigenvectors
corresponding to the normal modes.

\section{Quantum phase transitions}
\label{qpt}
In this section we consider the critical behavior of some of the continuous
quantum transitions among the phases in Fig~\ref{phasediag1}.
The transition between the paramagnet A and the N\'{e}el phase C
has already been discussed in some detail in Ref~\cite{CSY}. We will
present below a discussion of the transitions from the N\'{e}el (C)
and quantized ferromagnet (B) phases to the canted phase D.
There are also a number of multicritical points in Fig~\ref{phasediag1},
but we will not discuss them here.
As in previous sections, the discussion below will implicitly assume that
$d>1$, 
and we will obtain results for all such  $d$.

\subsection{N\'{e}el (C) to Canted (D) Transition}
\label{CtoD}

We begin with some general scaling ideas. 
This transition involves the onset of a mean value in a conserved charge, 
and this could possibly impose constraints on the critical exponents. In an
earlier
paper~\cite{conserve}, one of us had studied such constraints for the case
of a
transition from a paramagnet to a partially polarized ferromagnet. In the
present case,
the `paramagnet' is the N\'{e}el phase C, and the orientation of the N\'{e}el
order
parameter has As we show below, this is an important difference, which
modifies the
scaling relations.
 
The order parameter describing this transition is the coarse-grained
angular momentum
density
$\vec L_{\bot}(x, t)$ in a direction perpendicular to
the direction of N\'{e}el ordering which we 
choose to be the $z$ axis.  
The response of the system to a field that couples 
to $\vec L_{\bot}$ is described by the 
correlation function
\begin{equation}
\chi^{ab}(\vec k, \omega) = i\int d^d x dt~ \exp(-i(\vec k \cdot \vec x -
\omega t))  
  \langle [L^a (\vec x, t), L^b (0, 0)]\rangle \theta(t) = \delta^{ab}\chi(\vec
k,\omega),
\end{equation}
with $a,b = 1,2$, and 
where the angular brackets denote both thermal and quantum averaging.
For simplicity, we will only discuss here the scaling at zero temperature.
The behavior at finite temperature can be obtained from finite size
scaling through standard arguments\cite{fwgf,CSY}. On
approaching the transition from the the N\'{e}el phase, we expect that
$\chi$ satisfies the scaling form
\begin{equation}
\label{scal.form}
\chi(\vec k, \omega) = \frac{a_{0}}{k^{2 - \eta}}g(k\xi,
a_{1}\omega\xi^{z})
\end{equation}
where $\xi$ is the diverging correlation length associated
with the second order transition. $g(x, y)$ is a universal function
of it's variables and $a_0$ and $a_1$ are (possibly non-universal)
constants.
The susceptibility to a static, uniform external field that couples
to $\vec L_{\bot}$ is given by $\chi_0 = \lim_{k \rightarrow
0}\lim_{\omega \rightarrow
0} \chi(\vec k,\omega)$ and diverges with an exponent $\gamma = \nu(2 -
\eta)$.
It is possible to derive a scaling law relating $z$ and $\eta$,
if we make the additional assumption that the stiffness,
$\rho_s$, of the system to twists in the direction of the 
N\'{e}el order parameter has a finite non-zero value at the
transition. While we have no proof that this is so
in all cases, it is physically reasonable in view of the 
fact that the N\'{e}el order parameter is finite and non-vanishing
across the transition. The explicit computations below will obey this
assumption.

Throughout the N\'{e}el phase at very low $k, \omega$, 
$\chi(\vec k,\omega)$ has the form predicted by hydrodynamics:
\begin{equation}
\label{hydro}
\chi(\vec k,\omega) = \frac{\chi_{0}k^2}{ k^2 - \omega^2 /v_{s}^2}
\end{equation}
with the spin wave velocity $v_{s}$ given by
\begin{equation}
\label{swv}
v_{s}^2 = \frac{\rho_s}{\chi_0}
\end{equation}
The divergence of $\chi_0$ at the transition thus implies
that the spin-wave velocity vanishes.
In the vicinity of the critical point, we require that the scaling
form (\ref{scal.form}) reduce to the hydrodynamic
expression (\ref{hydro}) in the limit $k\xi \rightarrow 0,
a_1\omega \xi^z \rightarrow 0$. For this to happen, the function
$g(x, y)$ must satisfy
\begin{equation}
g(x, y) \rightarrow \frac{x^{4 - \eta}}{x^2 - y^2}~~~
as~~x,y \rightarrow 0
\end{equation}
This gives the scaling of $\chi_0$ and $v_{s}^2$ as
$\chi_0 \sim \xi^{2 - \eta}$, $v_{s}^2 \sim \xi^{2(1 - z)}$.
(The $\chi_0$ scaling is equivalent to the relation
$\gamma = \nu(2 - \eta)$.)
From the hydrodynamic expression (\ref{swv}), we then get
\begin{equation}
\label{zscal}
z = 2 - \frac{\eta}{2}
\end{equation}
Below we will verify this scaling relation by an
explicit renormalization group (RG) calculation. This scaling relation
replaces the
relation
$z=d+2-\eta$ proposed earlier~\cite{conserve} for the case of onset of
ferromagnetic
order from rotationally invariant paramagnetic phases; as we will see in
more detail
below, the latter relation has been modified here because $1/\rho_s$ (the
dimensionful
inverse of the stiffness of the N\'{e}el order) behaves like a ``dangerously
irrelevant
coupling''.

\subsubsection{Renormalization group analysis; ${\cal P}$ symmetry present}
\label{secrgmeq0}
We will examine here the behavior of the action (\ref{lsigma1}) for
systems with ${\cal P}$ symmetry under RG transformations. 
As noted earlier, this action undergoes a phase transition as a function of
the tuning
parameter $r_4$ from a N\'{e}el phase C
($r_4>0$ in mean-field theory) with $\langle \phi \rangle = 0$ to a canted
phase D ($r_4
< 0$ in mean-field theory) with $\langle \phi \rangle \neq 0$. The field
$\psi$
represents fluctuations of the N\'{e}el order about its mean value and we
always have
$\langle \psi \rangle = 0$.

Notice that the $\psi$ field plays a role analogous to the spin-wave modes
of the usual
$O(3)$ non-linear sigma model~\cite{nls}: their correlations are always
gapless, and no
mass-term for them is ever generated. It therefore seems natural at first
to construct an
RG under which the field $\psi$ remains dimensionless and its effective
stiffness
$K_2$ flows. So for the RG transformation
\begin{equation}
x^{\prime} = x/s~~~~~~~~\tau^{\prime} = \tau / s^{\bar{z}},
\end{equation}
where $\bar{z}$ is an unknown exponent, 
we have $\psi^{\prime} ( x^{\prime}, \tau^{\prime}  ) = \psi (x,\tau)$. At
tree
level invariance of the action requires
\begin{equation}
K_2^{\prime} = K_2 s^{d+\bar{z}-2}
\label{k2flow}
\end{equation}

In contrast, the field $\phi$ is a soft-spin order parameter, and should
behave under
RG like a conventional Landau-Ginzburg field. In this case, we fix $K_3 =
\mbox{constant}$ which requires
\begin{equation}
\phi^{\prime} = \phi s^{(d+\bar{z}-2)/2} 
\end{equation}
at tree level. Now demanding that the $K_1 ( \phi^{\ast} \partial_{\tau}
\psi -  \phi
\partial_{\tau} \psi^{\ast})$ be invariant we obtain, again at tree level
\begin{equation}
\bar{z} = d+ 2
\label{zbar}
\end{equation}
This appears to be the tree-level version of the identity $z=d+2-\eta$
discussed in
Ref~\cite{conserve} for the ferromagnetic onset transition, and it is
tempting
therefore to identify $\bar{z}$ as the dynamic critical exponent. Notice,
however that
for this value of $\bar{z}$, $K_2$ flows to infinity for all $d > 0$, and
there is thus
the possibility that the dimensionful constant $K_2^{-1}$ could appear in
determining
an important low-frequency scale which is not characterized by the
exponent $\bar{z}$. 
This is indeed what happens, as we can easily see by studying the
Gaussian
fluctuations at the critical point $r_4 = 0 $: the fluctuations involve
modes at 
a frequency $\omega \sim \sqrt{K_2 K_3 / K_1^2} k^2$ with momenta $k$.
This dispersion
is consistent with the scaling (\ref{k2flow}) of $K_2$ and the value
(\ref{zbar}) for
$\bar{z}$. However the present approach is quite awkward, and it is clear
that a scheme
with dynamic exponent $z=2$ would be preferable. 

Such a scheme is motivated by the realization that $K_2 \rightarrow
\infty$ in the
above RG (and so $1/K_2$ is dangerously irrelevant). As a large $K_2$
suppresses
fluctuations of $\psi$, we can safely integrate out the $\psi$ field
completely from
(\ref{lsigma1}), with the expectation that only the leading terms in
$1/K_2$ will be
important. Such a computation leads to the following effective action for
the $\phi$
field
\begin{equation}
{\cal S} = \int \frac{d^d k}{(2 \pi)^d}\frac{d \omega}{2 \pi}
\left (k^2 +
        b_0 \frac{\omega^2}{k^2} + r_0 \right) |\phi(\vec k, \omega)|^2
      + \frac{u}{2} \int d^d x d\tau  |\phi(\vec x,\tau)|^4 .
\label{actionmeq0}
\end{equation}
We have rescaled the field $\phi \rightarrow \phi/\sqrt{K_3}$, and
introduced the
couplings $b_0 = K_1^2 / (K_2 K_3)$, $r_0 =r_4 / K_3$ and $u = u_1 /
K_3^2$.
Notice now that there is a long-range term associated with the $\omega^2
/ k^2$
non-analyticity in the $\phi$ propagator. Standard methods to
analyze field theories with long-range couplings and anisotropic scaling
in spacetime
are available~\cite{bc}, and they can be applied to the action ${\cal S}$ in
(\ref{actionmeq0}) in a straightforward manner. The non-linearities in
${\cal S}$
are controlled by the dimensionless coupling constant
\begin{equation}
g_0 = u b_0^{-1/2} S_d \mu^{d-2},
\label{defg0}
\end{equation}
where $S_d = 2 \pi^{d/2} /(\Gamma(d/2) (2 \pi)^d )$ is a phase space factor
and $\mu$ is a momentum scale. The dependence on
$\mu$ indicates that the Gaussian
fixed point controls the infrared behavior for $d>2$: this fixed-point has
the exponents $z=2$, $\eta=0$, and $\nu = 1/2$. 

For $d<2$, a RG analysis using an
expansion in powers of $\epsilon = 2-d$ is necessary. As usual, we define
the
renormalized field  $\phi = \sqrt{Z} \phi_R$, and the renormalized
couplings
$g_0 = (Z_4 / Z^2) g$ and $b_0 = (Z_b / Z) b$. Note that the coupling $g$
here should
not be confused with the $g$ in the quantum rotor Hamiltonian
(\ref{hrotor}). The
renormalization constants
$Z$, $Z_4$, $Z_b$ are determined in a minimal subtraction scheme and we
obtained (some details appear in Appendix~\ref{apprg1}):
\begin{eqnarray}
Z &=& 1 - \frac{g^2}{72  \epsilon} + \ldots \nonumber \\
Z_4 &=& 1 + \frac{5g}{4 \epsilon} + \ldots \nonumber \\
Z_b &=& 1
\end{eqnarray}
The result for $Z_b$ is expected to be exact to all orders in $g$---there
are no
$\omega^2 / k^2$ terms generated in the two-point function in the loop
expansion.
The above results required the evaluation of certain one and two-loop
Feynman
graphs---these are most easily evaluated by writing each propagator in
spatial and time
co-ordinates and then doing the necessary integrals (Appendix~\ref{apprg1}).
 We can now obtain the $\beta$ function for
$g$:
\begin{equation}
\beta_g = \left. \mu \frac{\partial}{\partial \mu} g \right|_{B} = 
-\epsilon g + \frac{5 g^2 }{4}
\end{equation}
where the derivative is taken for a fixed bare theory. The $\beta$ function
has a fixed
point at $g = g^{\ast} = 4 \epsilon/5$, which is the infrared stable
fixed point
for $d<2$. Scaling exponents of the critical point can now be determined:
\begin{eqnarray}
\eta &=& \left. \mu \frac{\partial}{\partial \mu} \ln Z \right|_{B, g =
g^{\ast}}
 = \frac{4 \epsilon^2}{225} \nonumber \\
z &=& 2 + \frac{1}{2}\left. \mu \frac{\partial}{\partial \mu} \ln
\frac{Z_b}{Z}
\right|_{B, g = g^{\ast}}
 = 2 - \frac{2 \epsilon^2}{225} 
\label{zetavalue}
\end{eqnarray}
Note that the scaling relation (\ref{zscal}) is obeyed, and is a consequence
of $Z_b =
1$. The value of $r$ controls the deviation of the system from criticality,
and the
renormalization of $\phi^2$ insertions will determine the value of the
critical
exponent $\nu$. We define its renormalization by $r_0 = (Z_2 / Z) r$, and
obtain in a
similar manner
\begin{equation}
Z_2 = 1 + \frac{g}{2 \epsilon}
\end{equation}
which leads to 
\begin{equation} 
\frac{1}{\nu} = 2 - \left. \mu \frac{\partial}{\partial \mu} \ln
\frac{Z_2}{Z}
\right|_{B, g = g^{\ast}} = 2- \frac{2 \epsilon}{5}.
\label{nuvalue}
\end{equation}

\subsubsection{Renormalization group analysis; ${\cal P}$ symmetry absent}
\label{secrgmneq0}

As we noted earlier in Section~\ref{secmneq0}, in the absence of ${\cal
P}$ symmetry,
the most important modification of the action  (\ref{lsigma1}) is that a
term
$ K_5((\nabla \psi^\ast).(\nabla \phi) +
                          (\nabla \psi).(\nabla \phi^\ast))$ is now permitted.
As in Section~\ref{secrgmeq0} we may now integrate out $\psi$ from
such an action,
which leads to the following modified version of (\ref{actionmeq0}):
\begin{equation}
{\cal S} = \int \frac{d^d k}{(2 \pi)^d}\frac{d \omega}{2 \pi}
\left (k^2 +
        b_0 \frac{\omega^2}{k^2} - i \lambda_0 b_0^{1/2} \omega + r_0
\right) |\phi(\vec
k,
\omega)|^2
      + \frac{u}{2} \int d^d x d\tau  |\phi(\vec x,\tau)|^4 .
\label{actionmneq0}
\end{equation}
The new feature is the $i \omega$ term in the propagator. Associated with
this term is
the dimensionless coupling $\lambda_0$; power-counting shows that
$\lambda_0$ marginal at
tree level. Indeed $\lambda_0$ remains marginal at one-loop, and a two-
loop calculation
is necessary to decide if the $\lambda_0 = 0$ fixed point of
Section~\ref{secrgmeq0} is
stable. 

We define the renormalization of $\lambda$ by $\lambda_0 b_0^{1/2} 
= (Z_\lambda / Z) \lambda b^{1/2}$. The new values of the renormalization
constants
can now be computed to be (Appendix~\ref{apprg1}):
\begin{eqnarray}
Z &=& 1 - \frac{g^2}{2 \epsilon} \frac{ \lambda^2 + 9}{(\lambda^2 +
4)
(2 \lambda^2 + 9)^2} + \ldots
\nonumber
\\ Z_4 &=& 1 + \frac{g}{2 \epsilon} \frac{\lambda^2 + 20}{(\lambda^2
+ 4)^{3/2}} +
\ldots
\nonumber \\ 
Z_b &=& 1 \nonumber \\
 Z_2 &=& 1 + \frac{4g}{ \epsilon} \frac{1}{(\lambda^2 + 4)^{3/2}} +
\ldots
\nonumber \\
Z_{\lambda} &=& 1 - \frac{g^2}{2 \epsilon} \frac{1}{(\lambda^2 + 4)
(2 \lambda^2 + 9)} + \ldots
\label{Zvalues}
\end{eqnarray}
These results require an expansion in powers of $g$, but the dependence on
$\lambda$ is
exact at each order.
The $\beta$ functions of $g$ and $\lambda$ can now be determined
\begin{eqnarray}
\beta_g &=& \left. \mu \frac{\partial}{\partial \mu} g \right|_{B} = 
-\epsilon g + \frac{g^2 }{2} \frac{\lambda^2 + 20}{(\lambda^2 +
4)^{3/2}} \nonumber
\\
\beta_\lambda &=& \left. \mu \frac{\partial}{\partial \mu} \lambda
\right|_{B} = 
- \frac{g^2 }{2 } \frac{  \lambda (5\lambda^2 + 27)}{(\lambda^2 + 4)
(2 \lambda^2 + 9)^2}
\label{betafuncs}
\end{eqnarray}
One of the fixed points of (\ref{betafuncs}) is the model with ${\cal P}$
symmetry of
Section~\ref{secrgmeq0}: 
\begin{equation}
\lambda^{\ast} = 0$,~~~ $g^{\ast} = 4 \epsilon / 5.
\end{equation}
However, this fixed point is {\em unstable} in the infrared; the stable
fixed point
is instead 
\begin{equation}
\lambda^{\ast} = \infty$,~~~ $(g/\lambda)^{\ast} = 2 \epsilon ;
\end{equation} 
for $\lambda \rightarrow \infty$, the loop corrections become functions
of
$g/\lambda$, which is therefore the appropriate measure of the strength
of the
non-linearities. Near this stable fixed point, the frequency dependence of
the
propagator is dominated by the
$i \omega$ term and the $\omega^2 / k^2$ term can be neglected. The
exponents at this
fixed point can be determined as in (\ref{zetavalue}) and (\ref{nuvalue}) 
from
(\ref{Zvalues}) and we find the same exponents as those of the $d>2$
Gaussian fixed
point:
\begin{equation}
z = 2 ~~~~~~~\eta=0~~~~~~\nu=1/2
\end{equation}
Although these exponents are the same for $d>2$ and for $d<2$, it is
important to note
that the $d<2$ theory is {\em not} Gaussian---there is a finite fixed point
interaction
$(g/\lambda)^{\ast} = 2 \epsilon$ which will make the scaling
functions and finite
temperature properties different from those of the Gaussian theory. The
structure of
this non-Gaussian fixed point has been explored earlier~\cite{fwgf,sss} in
in a different context and more details are provided in Appendix~\ref{appcalc}). 
Also note that the flow of $\lambda$ to this fixed point is slow,
and will lead to logarithmic corrections to scaling.

\subsection{Quantized Ferromagnet (B) to Canted (D) Transition}
\label{BtoD}
The ferromagnetic order parameter $\langle \hat{L}_{\mu} \rangle$
is non-zero in both the B and D phase, while the component of
$\hat{n}_{\mu}$
orthogonal to $\langle \hat{L}_{\mu} \rangle$ plays the role of the order
parameter. The roles of $\hat{n}_{\mu}$ and $\hat{L}_{\mu}$ are therefore
reversed
from the considerations in Sections~\ref{struct}
and~\ref{CtoD}, and the analysis here will, in some sense, be dual
to the earlier
analyses. In the following we will assume that $M\neq 0$ and that ${\cal
P}$ is
not a good symmetry; the special properties of the $M=0$ case will be
noted as
asides.

To begin, we need a rotationally-invariant field theory which describes
the
dynamics
in one of the quantized B phases and across its phase boundary to the D
phase.
Let us work in the $\rm{B}_{\ell_0}$ phase {\em i.e.} the phase in which
the
quantized moment is $\ell_0 \neq 0$ per rotor. If we now proceed with a
derivation of the path integral as in Section~\ref{struct}, we find that the
single site functional integral (\ref{zloc}) has to be evaluated in a regime
in
which (\ref{eloc}) has a minimum at $\ell = \ell_0$: this is quite difficult
to
do. We circumvent this difficulty by the following stratagem. We
consider a somewhat different model, which nevertheless has a quantized
ferromagnetic phase (with a moment ${\ell_0}$ per site) and a canted
phase
separated by a second order transition; this model has an enlarged Hilbert
space, but the additional states have a finite energy, and we therefore
expect
that its transition is in the same universality class as in the
original rotor model. The new model is an ordinary Heisenberg ferromagnet
coupled
to  a quantum rotor model through short-ranged interactions. The
Hamiltonian
$H_{new} = H_{Heis} + H_{rot} + H_{int}$ where
\begin{eqnarray}
\label{Heis_rot}
H_{Heis}& = & -\Delta \sum_{\langle i,j \rangle} \hat S_{i\mu} \hat
S_{j\mu}
\nonumber \\ H_{rot} & = & g\sum_i \hat L_{i\mu}^2 - J\sum_{\langle i,j
\rangle}
                          \hat n_{i\mu} \hat n_{j\mu} \nonumber \\
H_{int} & = & -(\sum_{i,j} K_{i-j}\hat S_{i\mu} \hat L_{i\mu} +
                \sum_{i,j} M_{i-j} \hat S_{i\mu} \hat n_{i\mu})
\end{eqnarray}
Here $\hat S_{i\mu}$ is an ordinary Heisenberg spin operator acting on a 
spin $\ell_0$ representation, and it commutes with the $\hat n_{\mu}$
and 
$\hat L_{\mu}$. The $K_{i-j}$ and $M_{i-j}$ are some short-ranged
interactions
and $\Delta$ is a positive constant.
(There is an overlap in the symbols for the couplings above with
those used before;
 this is simply to prevent a proliferation
of symbols,
and it is understood that there is no relationship between the values in
the two cases.) The total conserved angular momentum is
$\sum_i L_{i\mu} +
\hat S_{i\mu}$.
The Hilbert space on each site consists of the spin ${\ell_0}$ states of
the
Heisenberg spin and the $\ell = 0,1,2,3,\ldots$ states of each rotor.
Ignoring
inter-site couplings, the lowest energy states on each site are a multiplet
of
angular momentum $\ell_0$, similar to the $\rm{B}_{\ell_0}$ phase of
the
original model.

It is now easy to see that $H_{new}$ exhibits a quantized 
ferromagnetic phase (with a moment ${\ell_0}$ per site) and a canted
phase for
appropriate choices of  parameters. First imagine that all the $K_{i-j}$'s
and
$M_{i-j}$'s are  zero and that $g\gg J$. Then naive perturbation theory in
$J/g$
tells us that the ground state is a quantized ferromagnet with total
angular
momentum
$\ell_0$ per site. Now imagine introducing non-zero values of the $K$ and
$M$
couplings. The rotor variables then see an effective ``magnetic field'' that
couples to the 
$L_{i\mu}$, and an effective field that couples to the $\hat n_{i\mu}$.
The latter has the effect of introducing an expectation value for
the component of $\hat n_{\mu}$ parallel to $\langle \hat S_{\mu}
\rangle$.
 The ``magnetic field'' does not introduce an
 expectation value of $L_{\mu}$ if it is weak.
As the strength of the ``magnetic field'' is increased by varying the $K_{i-
j}$'s,
it induces a transition into a phase with a non-zero, non-integer
expectation
value of $\L_{\mu}$, and a non-zero expectation value for the component
of $\hat n_{\mu}$ perpendicular to $\langle \hat S_{\mu} \rangle$. 
This is the transition to the canted phase.

The derivation of the path integral of this model is easy;
 because the Hilbert space has now been expanded,
we have
also have to introduce coherent states of spin ${\ell}_0$ at each time
slice in the Trotter product. The final functional integral will be
expressed in terms
of the field $n_{\mu}$ of the quantum rotor, and a unit
vector field
$s_{\mu}$ ($s_\mu^2 = 1$) representing the orientation of the spin
$\hat{S}_{\mu}$.
 Finally, as in
Section~\ref{struct}, we also impose the constraint $n_{\mu} s_{\mu} =
c$
to project out fluctuations which have a local energy cost (this cost is
due to
the effective field, noted above, that acts between the $s_{\mu}$ and
$n_{\mu}$).
Our final form of the modified sigma model appropriate for the transition
from a
quantized ferromagnet
$B_{\ell_0}$ to the canted phase $D$ is
\begin{displaymath}
Z = \int {\cal D} n_{\mu} {\cal D} s_{\mu} \delta \left( s_\mu^2 - 1 \right)
\delta \left( n_\mu s_\mu - c \right)
\exp\left( - \int d^d x
\int_0^{\beta} d
\tau {\cal L} \right)
\end{displaymath}
\begin{eqnarray}
{\cal L} = && i M_0 A_{\mu} (s) \frac{\partial s_\mu}{\partial \tau}
- M_0 H_{\mu} s_{\mu} +
\frac{K_1}{2} \left( \frac{\partial n_{\mu}}{\partial
\tau} + i \epsilon_{\mu\nu\lambda} ( \rho s_{\nu} - H_{\nu}) n_{\lambda}
\right)^2
\nonumber \\
&&~~~~~~~+ \frac{K_2}{2} (\nabla n_{\mu})^2
+ \frac{K_3}{2} (\nabla s_{\mu})^2 + K_4 (\nabla n_{\mu}) (\nabla
s_{\mu})
+
\frac{r_1}{2} n_{\mu}^2  + \frac{u_1}{8} (n_{\mu}^2)^2
\label{sigmaBtoD}
\end{eqnarray}
Again there is an overlap in the symbols for the couplings above with
those used in
Secs~\ref{struct} and~\ref{BtoD},
and it is understood that there is no relationship between the values in
the two cases. There is no length constraint on the components of
$n_{\mu}$ orthogonal to
$s_{\mu}$, which act as a soft-spin order parameter for the transition
from phase B to
phase D. The
$\rho s_\nu$ term is the effective internal ``magnetic field'' (noted
above) that the
Heisenberg spins impose on the
$n_{\mu}$, with
$\rho$ a coupling constant that measures its strength.
Note also that
\begin{equation}
 K_4 = c=0~~~\mbox{if
${\cal P}$ is a symmetry.}
\end{equation}
The coupling $M_0 = \l_0 a^{-d}$, where $a^d$ is the volume per rotor,
and, as before, $H_\mu$ is the external magnetic field. The first term in
${\cal L}$ is the Berry phase associated with quantum fluctuations of the
ferromagnetic
moment: $A_{\mu} (s)$ is the vector potential of a unit Dirac monopole at
the
origin of $s$ space, and is defined up to gauge transformations by
$\epsilon_{\mu\nu\lambda} \partial A_{\lambda}/\partial s_{\nu} =
s_\mu$.

The onset of phase D is signalled by a mean value in $n_\mu$ orthogonal to
the average
direction of $s_\mu$. Using a reasoning similar to that above
(\ref{paramet}),
we parametrize the fields of the modified sigma model as
\begin{eqnarray}
\vec{s} &=& \left( \frac{\phi + \phi^\ast}{2} (2 - |\phi|^2)^{1/2}, \frac{\phi-
\phi^\ast}{2i} (2 - |\phi|^2)^{1/2}, 1 -  |\phi|^2 \right) \nonumber \\
\vec{n} &=&  c \vec{s} +
\left( \frac{\psi + \psi^\ast}{\sqrt{2}}, \frac{\psi -
\psi^\ast}{\sqrt{2}i},
-\frac{(\psi^\ast \phi + \phi^\ast \psi)(2- |\phi|^2)^{1/2}}{ \sqrt{2} ( 1
-  |\phi|^2
)}
\right).
\label{paramet2}
\end{eqnarray}
We have chosen a non-standard parametrization of the unit vector field
$s_\mu$ in terms of a complex scalar $\phi$: this choice is a functional
integral
version of the Holstein-Primakoff decomposition of spin operators, and
ensures
that the Berry phase term
$i M_0 A_{\mu} (s) (\partial s_\mu/\partial \tau)$ is exactly equal to
$M_0 \phi^{\ast} (\partial \phi/\partial \tau)$. The parametrization of
$n_{\mu}$
then follows as in (\ref{paramet}) and ensures that $n_\mu s_\mu = c$.
The two complex fields $\psi, \phi$, correspond to the 4 degrees of
freedom of the hybrid sigma model. The field $\phi$ represents
oscillations
of the $\vec{s}$ about its mean value: by definition,
we will have $\langle \phi \rangle = 0$ in both the B and D phases.
The field $\psi$ measures
the amplitude of $n_{\mu}$ orthogonal to the instantaneous value of
$s_{\mu}$: we
have
$\langle \psi \rangle =0$ in the quantized ferromagnet B, while
$\langle \psi \rangle \neq 0$ in the canted phase D. We now insert
(\ref{paramet2})
into (\ref{sigmaBtoD}) and obtain
\begin{eqnarray}
&&~~~~~~~~~Z_{\sigma} = \int {\cal D} \psi {\cal D} \psi^\ast \frac{
{\cal D}
\phi  {\cal D} \phi^\ast}{1 -  |\phi|^2} \exp \left( - \int d^d x \int_0^{\beta}
d \tau ( {\cal L}_{\sigma} + \cdots ) \right) \nonumber \\
&&~~~~~~~~~~~~~{\cal L}_{\sigma } =
M_0 \phi^\ast \frac{\partial \phi}{\partial \tau} + K_5 \psi^{\ast}
\frac{\partial \psi}{\partial \tau} + K_6 \left(
\psi^{\ast} \frac{\partial \phi}{\partial \tau} + \phi^{\ast} \frac{\partial
\psi}{\partial
\tau} \right) \nonumber \\
&&~~~~~~~~~~~~~~~~~~~~~~+ H_z \left( - M_0 +  M_0 |\phi|^2 + K_5
|\psi|^2 + K_6 (\psi^\ast
\phi + \psi \phi^{\ast} ) \right) \nonumber\\
&&~~~~~~~~~~~~~~~~~~~~~~+ K_2 |\nabla \psi|^2 + K_7 |\nabla \phi|^2 +
K_8 (\nabla \psi^\ast
\nabla \phi + \nabla \psi \nabla \phi^\ast ) \nonumber \\
&&~~~~~~~~~~~~~~~~~~~~~~+ r_2 \left( |\psi|^2 + \frac{1}{2} (\psi^\ast
\phi + \psi \phi^{\ast}
)^2 \right)
 + \frac{u_1}{2} |\psi|^4 
\label{lsigma2}
\end{eqnarray}
where $K_5 = 2 \rho K_1 $, $K_6 = c \rho K_1$, $K_7 = K_3 + 2 c K_4 + c^2
K_2$,
$K_8 = K_4 + c K_2$,
$r_2 = r_1  - K_1 \rho^2 + u_1 c^2 /2 $. Stability of the action requires
that $K_2 > 0$, $K_7 > 0$, and $K_2 K_7 > K_8^2$; these conditions will be
implicitly assumed
below, and are the analog of the requirement in the mean-field theory of
Section~\ref{mft} that $M$ not be too large. Without loss of generality,
we can assume that $M_0 > 0$; however the signs of $K_5$ and 
$M_0 K_5 - K_6^2$ can be arbitrary, and these will play an important role in
our analysis below. Systems with 
$M=0$, and the symmetry
${\cal P}$, will have $K_6 = K_8 = 0$. We
have explicitly displayed only terms up to quartic order in the fields.
Anticipating the RG analysis below, we have also dropped irrelevant terms
with additional time derivatives.
The field $H_{\mu} = (0,0,H_z)$ has been assumed to point along the $z$
direction.

Notice that the coupling to $H_z$ could have been deduced by the gauge
invariance
principle~\cite{conserve} which demands the replacement
$\partial/\partial \tau 
\rightarrow  \partial/\partial \tau + H_z$ ($\partial/\partial \tau 
\rightarrow  \partial/\partial \tau - H_z$) when acting on the $\phi$,
$\psi$
($\phi^{\ast}$, $\psi^{\ast}$) fields. The magnetization density, $\langle 
\cal M \rangle$ 
is given by 
\begin{equation}
\langle 
{\cal M} \rangle = 
- \frac{\partial {\cal F}}{\partial H_z},
\label{defcalm}
\end{equation}
where ${\cal F}$ is the
free energy
density. The gauge invariance of the action ensures that at $T=0$, 
this magnetization density is exactly $M_0$, as long as $\langle \psi
\rangle = 0$
{\em i.e.\/} we are in phase B.

We can easily do a small fluctuation, normal mode, analysis  in both
phases of
(\ref{lsigma2}). In the quantized ferromagnetic phase B ($r_2 > 0$) we
have the usual
ferromagnetic spin waves with $\omega = K_7 k^2 / M_0$ at small $k$ and
also a massive mode
with
$\omega \sim r_2 / (K_5 - K_6^2 / M_0 )$ (we have assumed here and in
the remainder of the
paragraph that $H_z = 0$). The massive mode becomes gapless at
$r_2 = 0$, signaling the onset of the canted phase D. In this canted phase,
we
have $\langle \psi \rangle = \sqrt{-r_2 / u_1}$. Analysis of fluctuations
about
this mean value for $\psi$ gives two eigenfrequencies which vanish as $k
\rightarrow 0$;
the first has a linear dispersion
\begin{equation}
\omega^2 = \frac{ 2 |r_2| W}{
(M_0 K_5 - K_6^2)^2} k^2,
\end{equation}
while the second disperses quadratically
\begin{equation}
\omega^2 = \frac{ K_7 ( K_2 K_7 - K_8^2)}{W} k^4,
\end{equation}
where $W = (M_0 \sqrt{K_2} - K_6 \sqrt{K_7})^2 + 2 K_6 M_0
(\sqrt{K_2 K_7} - K_8) > 0$.
These results for phase D are in agreement with those obtained earlier
using a `dual'
approach with the action (\ref{lsigma1}) in Sections~\ref{secmeq0}
and~\ref{secmneq0}.

It is also useful to compare and contrast the action (\ref{lsigma1}) for
the
transition
from C to D with the action (\ref{lsigma2}) above for the B to D transition.
Notice that the roles of the fields are exactly reversed--in the former
$\phi$ was
the order parameter for the transition while $\psi$ described spectator
modes
which were required by symmetry to be ``massless''; in the latter
$\psi$ is the order parameter while $\phi$ is ``massless'' spin-wave mode.
The two models however differ significantly in the nature of the time
derivative
terms in the action. In (\ref{lsigma2}) there are terms linear
in time derivatives for both the $\psi$ and $\phi$ fields, while there are
no such
terms in (\ref{lsigma1}) (the time derivative term
involving an off-diagonal $\phi$, $\psi$ coupling is present in both cases,
though).
As a result the spectator spin-wave $\phi$ mode in (\ref{lsigma2}) has
$\omega \sim
k^2$, while the spectator spin-wave $\psi$ mode in (\ref{lsigma1}) has
$\omega \sim k$.
These differences have significant consequences for the RG analysis of
(\ref{lsigma2}), which will be discussed now.

\subsubsection{Renormalization group analysis}
\label{secrgBtoD}
The logic of the analysis is very similar to the `dual' analysis in
Sections~\ref{secrgmeq0}
and~\ref{secrgmneq0}. We expect the stiffness of the spectator $\phi$
modes 
flows to infinity, and hence it is valid to simply integrate out the $\phi$
fluctuations. This gives us the following effective action for the $\psi$
field
(which, recall, measures the value of $\hat{n}_\mu$ in the direction
orthogonal
to the ferromagnetic moment):
\begin{eqnarray}
{\cal S} = && \int \frac{d^d k}{(2 \pi)^d}\frac{d \omega}{2 \pi}
\left (-i K_5 \omega  + K_2 k^2  - \frac{(-iK_6 \omega + K_8 k^2)^2}{-
i M_0 \omega
+ K_7 k^2} + r_2 \right) |\psi(\vec k,
\omega)|^2
      \nonumber \\
&&~~~~~~~~~~~~~~~~~~~~~~~~~~~~~~~~~~~~~~~~
+ \frac{u_1}{2} \int d^d x d\tau  |\psi(\vec
x,\tau)|^4 .
\label{actionbtod}
\end{eqnarray}
We have written the action for $H_z = 0$, and the $H_z$ dependence can be
deduced
from the mapping
$-i\omega \rightarrow -i\omega + H_z$. Simple power-counting
arguments do not permit any
further simplifications to be made to this action. Despite the apparent
formidable
complexity of its form, the RG properties of (\ref{actionbtod}) are quite
simple, as we shall
now describe.

We will consider the cases with and without ${\cal P}$ symmetry separately.

\subsubsection*{{\em a.} ${\cal P}$ symmetry present}

Now we have $K_6 = K_8 = 0$. The non-analytic terms in the propagator
of (\ref{actionbtod}) disappear, and the remaining terms are identical to those
in the action of a dilute Bose gas with a repulsive interaction $u_1$. The
order parameter, $\psi$, and the spectator modes, $\phi$, are essentially
independent, with all couplings between them contributing only ``irrelevant''
corrections to the leading critical behavior.

The critical theory of the quantum transition in the dilute Bose gas has been
studied earlier~\cite{fwgf,sss} (see Appendix~\ref{appcalc}), and also appeared as the
fixed point theory in Section~\ref{secrgmneq0}. 
A measure of the strength of the loop corrections is the
dimensionless
coupling constant $g_0$ (the analog here of Eqn (\ref{defg0}))
\begin{equation}
g_0 = u_1 K_2^{-1} |K_5|^{-1} S_d \mu^{d-2}
\label{defg0a}
\end{equation}
(where, as before, $\mu$ is a renormalization momentum scale, $S_d$ is a
phase space factor, and the symbol $g$ is not to be confused with the coupling
$g$ in (\ref{hrotor})), which determines the scattering amplitude of two
pre-existing excitations. This amplitude undergoes a one-loop renormalization due
to  diagrams associated with repeated scattering of the two excitations, a process
which leads to the $\beta$ function
\begin{equation}
\beta_g = -\epsilon g +  \frac{g^2}{2}
\label{betadilute}
\end{equation}
where $\epsilon = 2-d$ and we have used the minimal subtraction method to define
the renormalized $g$ (this
$\beta$ function is in fact exact to all orders in
$g$---see Appendix~\ref{appcalc}). 
The infrared stable fixed point is $g^{\ast}=0$ for
$d>2$, and 
$g^{\ast} = 2 \epsilon$ for $d<2$. The exponents take the same values as in
Section~\ref{secrgmneq0} 
\begin{equation}
z = 2 ~~~~~~~\eta=0~~~~~~\nu=1/2.
\label{bosegasexponents}
\end{equation}
for all values of $d$. It is important to note, however, that the present
subsection and Section~\ref{secrgmneq0} have very different interpretations of
the order parameter. In Section~\ref{secrgmneq0} the exponent $\eta$ is
associated with the field $\phi$ which is proportional to the ferromagnetic
moment. Here the $\eta$ refers to the the field $\psi$, and the ferromagnetic
order parameter scales as $|\psi|^2$ as we discuss below.

The magnetization density order
parameter ${\cal M} (x, \tau
)$, is defined by generalizing (\ref{defcalm}) a space-time dependent external
magnetic field $H_z ( x, \tau)$. In the quantized ferromagnetic phase B,  clearly
$\langle {\cal M} \rangle =
M_0$. Therefore, the {\em deviation} $M_0 -  {\cal M}$ can serve as
another order
parameter for the B to D transition. Indeed,  this order parameter scales
as
$|\psi |^2$, and we can define an associated ``anomalous'' dimension
$\eta_{{\cal M}}$. A
simple calculation shows that $\eta_{{\cal M}} = d$, and so the
identity~\cite{conserve}
\begin{equation}
z = d + 2 - \eta_{{\cal M}}
\end{equation}
is always satisfied. As in Ref~\cite{sss}, we can also compute the
behavior
of the mean value of this conserved order parameter in the canted phase D.
For $d<2$, this behavior is universal and is given by
\begin{equation}
\langle {\cal M} \rangle = M_0 - \mbox{sgn}(K_5) \theta(-r_2) {\cal C}_d \left(\frac{|
r_2|}{K_2}
\right)^{d/2},
\label{calmval}
\end{equation}
where ${\cal C}_d$ is a universal number; we compute ${\cal C}_d$ in an
expansion in $\epsilon$ in Appendix~\ref{appcalc}.
Note that the magnetization density
can increase or decrease into the ferromagnetic phase, depending upon the sign
of $K_5$; this behavior is consistent with that of the mean field theory in
Section~\ref{mft}. Note also that (\ref{calmval}) is independent of the magnitude
of $K_5$. The magnitude of
$K_5$ however does determine the width of the critical region within which
(\ref{calmval}) is valid. For $|K_5|$ smaller than $\sqrt{r_2}$, it becomes
necessary to include the leading irrelevant frequency dependence in the
propagator of (\ref{actionbtod}) - a $\omega^2 |\psi (\vec{k}, \omega )|^2$ term.  
The point $K_5 = 0$ is a special multicritical point at which this new term is
important at all values of $r_2$; this multicritical point has $z=1$~\cite{fwgf},
and has $\langle {\cal M} \rangle$ continue to equal $M_0$ (to leading order) in
phase D.

\subsubsection*{{\em b. } ${\cal P}$ symmetry absent}

The theory without ${\cal P}$ symmetry is somewhat more involved, and it is
convenient to work with a more compact notation. We rescale $\psi(\vec{k}, \omega)
\rightarrow (M_0/\sqrt{W K_7}) \psi(\vec{k}, \omega)$, $\omega \rightarrow \omega
K_7$, and manipulate (\ref{actionbtod}) into the form
\begin{eqnarray}
{\cal S} = && \int \frac{d^d k}{(2 \pi)^d}\frac{d \omega}{2 \pi}
\left (-i\lambda_0 M_0 \omega  +  k^2  - b_0 \frac{k^4}{-
i M_0 \omega
+ k^2} + r_0 \right) |\psi(\vec k,
\omega)|^2
      \nonumber \\
&&~~~~~~~~~~~~~~~~~~~~~~~~~~~~~~~~~~~~~~~~
+ \frac{u}{2} \int d^d x d\tau  |\psi(\vec
x,\tau)|^4 ,
\label{actionbtodn}
\end{eqnarray}
where $r_0 = M_0^2 r_2 / W$, $u = u_1 K_7 M_0^4 /W^2$, $\lambda_0 = K_7
(M_0 K_5 - K_6^2 )/W$,
 and $b_0 = (K_6 K_7 - K_8 M_0)^2 / W K_7$. 
Notice that we have introduced two dimensionless couplings, $\lambda_0$ and
$b_0$, while $M_0$ has the dimensions of time/(length)${}^2$. The stability
conditions discussed below (\ref{lsigma2}) imply that $0 \leq b_0 < 1$. 

We now discuss the renormalization of the theory (\ref{actionbtodn}) as 
a perturbation theory in the dimensionless coupling constant
\begin{equation}
g_0 = u  S_d \mu^{d-2} /(M_0 |\lambda_0|),
\end{equation}
which is the analog of (\ref{defg0a}).
A key property of this perturbation theory, similar to that noted in
Section~\ref{secrgmeq0}, is that no terms sensitive to the ratio
$\omega /k^2$, as $\omega \rightarrow 0$, $k \rightarrow 0$, are ever
generated. As a result the term multiplying $b_0$ in (\ref{actionbtodn}) does
not undergo any direct renormalization. With this in mind, we define the
renormalized field by $\psi = \sqrt{Z} \psi_R$, and the renormalized couplings
by $\lambda_{0} = (Z_{\lambda} / Z) \lambda$, $b_0 = (1/Z) b$, $r_0 = (Z_2 /Z)
r$ and
$g_0 = (Z_4 / Z^2 ) g$. It remains to compute the renormalization constants $Z$,
$Z_{\lambda}$, $Z_2$, and $Z_4$.

In the perturbation theory in $g$, the sign of $\lambda$ (which is also the
sign of $M_0 K_5 - K_6^2$) plays a crucial role.  The physical interpretation of
this sign is the following: when $\lambda > 0$, the spin wave quanta, $\phi$,
and the order parameter quanta, $\psi$, carry the same magnetic moment; however
for $\lambda < 0$ they carry opposite magnetic moments. This was already clear
in the $K_6=0$ result (\ref{calmval}), where the condensation of $\psi$ lead to a
decrease or increase in the mean magnetization density depending upon the sign
of $K_5$. For $\lambda < 0 $, it is then possible to have a low-lying excitation of
a $\psi$ and a $\phi$ quantum, which has the same spin as the ground state. 
For the system without ${\cal P}$ symmetry, such an excitation will mix with the
ground state, and leads to some interesting structure in the renormalization
group.
In the actual computation, the sign of $\lambda$ determines the
location of the poles of the propagator of
(\ref{actionbtodn}) in the complex frequency plane. A simple
calculation shows that the propagator has two poles, and both poles
are in the same half-plane for $\lambda > 0$, while for $\lambda < 0$ the poles
lie on opposite sides of the real frequency axis.
We will consider these two cases separately below:

\noindent
\underline{$\lambda > 0$}\\ 
For the case where the poles are in the same half-plane, we can close the
contour of frequency integrals in the other half-plane, and as a result many
Feynman diagrams are exactly zero. In particular, all graphs in the expansion
of the self energy vanish. As a result, we have
\begin{equation} 
Z = Z_{\lambda} = Z_2 = 1
\end{equation}
There is still a non-trivial contribution to $Z_4$ (from the
`particle-particle' graph):
\begin{equation}
Z_4 = 1 + \frac{g}{2\epsilon} \frac{(\lambda + 1 - b)}{(1-b)(1+\lambda)}
\end{equation}
where, as before $\epsilon = 2-d$. These renormalization constants lead to
the
$\beta$ functions 
\begin{eqnarray}
\beta_b &=& 0 \nonumber \\
\beta_\lambda &=& 0 \nonumber \\
\beta_g &=& -\epsilon g + \frac{g^2}{2}\frac{(1+ \lambda -
b)}{(1-b)(1+\lambda)}
\end{eqnarray}
The vanishing of $\beta_b$ and $\beta_\lambda$ is expected to hold to all orders
in $g$; as a result, $b$ and $\lambda$ are dimensionless constants which can
modify the scaling properties. The exponents are however independent of the values
of $b$ and $\lambda$, and are identical to those of the system with ${\cal P}$
symmetry, discussed above in Section~\ref{secrgBtoD}a. 
The upper critical dimension is $2$, and for $d<2$ the magnetization density obeys
an equation similar to (\ref{calmval})
\begin{equation}
\langle {\cal M} \rangle = M_0 - \theta(-r_0) \bar{\cal C}_d | r_0|^{d/2}.
\label{calmval1}
\end{equation}
The `universal' number $\bar{\cal C}_d$ now does
depend upon
$b$ and
$\lambda$: the leading term in $\bar{\cal C}_d$ can be computed by
using the gauge-invariance argument to deduce the modification of
(\ref{actionbtodn}) in a field, using (\ref{defcalm})
to determine the expression for ${\cal M}$~\cite{limits}, determining the
fixed point of
$\beta_g$, and then using the method of Ref~\cite{sss}---this gives
\begin{equation}
\bar{\cal C}_d  = \frac{S_d}{2 \epsilon} \left( 1 + \frac{b}{\lambda} \right)
\frac{(1-b)(1+\lambda)}{1 + \lambda - b}
\label{rescalcd1}
\end{equation}
This result should be compared with the first term in (\ref{rescalcd}), which is
the value of ${\cal C}_d$ for the Bose gas. Terms higher order in $\epsilon$ are
also expected to be modified by $b$ and $\lambda$, but have not been computed.
The results (\ref{calmval1}), (\ref{rescalcd1}) are not useful for too small a value
of $\lambda$: the reasons are the same as those discussed below (\ref{calmval})
but with $\lambda$ playing the role of $K_5$.

\noindent
\underline{$\lambda < 0$}\\ 
The computations for $\lambda <0 $ are considerably more
involved and are summarized in Appendix~\ref{apprg2}. From (\ref{apprg2res1}),
(\ref{apprg2res1a})
and (\ref{apprg2res1b})
we can deduce the renormalization constants
\begin{eqnarray}
Z &=& 1 - \frac{g^2}{2\epsilon} R_1 (b,\lambda) \nonumber \\
Z_{\lambda} &=& 1 - \frac{g^2}{2\epsilon\lambda} R_2 (b,\lambda) \nonumber \\
Z_4 &=& 1 + \frac{g}{2\epsilon} R_3 (b, \lambda) \nonumber \\
Z_2 &=& 1 + \frac{g}{\epsilon} R_4 (b, \lambda),
\end{eqnarray}
where $R_1$, $R_2$, $R_3$, $R_4$ are functions of $b$ and $\lambda$ defined in
(\ref{apprg2res2},\ref{apprg2res3}). These now lead to the $\beta$ functions
\begin{eqnarray}
\beta_b &=& g^2 b R_1( b, \lambda)  \nonumber \\
\beta_\lambda &=& g^2 (\lambda R_1 (b,\lambda) - R_2 (b, \lambda) ) \nonumber \\
\beta_g &=& -\epsilon g + g^2 R_3 (b, \lambda)/2.
\label{nuts}
\end{eqnarray}
It is not difficult to deduce the consequences of these flows by some numerical
analysis aided by asymptotic analytical computations.

For $d \geq 2$ ($\epsilon < 0$) we  have 
$b(\mu) \rightarrow b^{\ast}$, $\lambda(\mu) \rightarrow \lambda^{\ast}$,
$g(\mu) \rightarrow 0$ in the infrared ($\mu \rightarrow 0$). 
Here $0 < b^{\ast} <1 $ and $\lambda^{\ast} < 0$, but the values of $b^{\ast}$,
$\lambda^{\ast}$ are otherwise arbitrary and determined by the initial
conditions of the flow; they acquire only a finite renormalization from their
initial vales.  The flow to this final state has a power-law dependence on $\mu$
for $\epsilon <0$, while it behaves likes $\sim 1/\log(1/\mu)$ for $\epsilon =
0$. The properties of the final state have only minor differences from 
$\lambda > 0$ case discussed above, and we will not elaborate on them.

The possible behaviors are somewhat richer for $\epsilon > 0$.
The flow in the infrared is either to a fixed line or a fixed point.\\
({\em i}) The fixed line is $2/3 < b^{\ast} < 1$, $\lambda^{\ast} = 0$,
$g^{\ast} = 2 \epsilon$; the flow of $g$ to $g^{\ast}$ is a power-law
in $\mu$, while that of $b$ and $\lambda$ is logarithmic:
\begin{equation}
\lambda (\mu) 
= - \frac{(1 - b^{\ast})}{2 \epsilon^2 b^{\ast} (3 b^{\ast} - 2) \log(1/\mu)}
~~,~~b(\mu) - b^{\ast} = \frac{(1-b^{\ast})^2}{4 \epsilon^2 (3 b^{\ast} - 2)^2
\log(1/\mu)}.
\end{equation}
({\em ii}) The fixed point is $b^{\ast} = 0$, $\lambda^{\ast} = - \infty$,
$g^{\ast} = 2 \epsilon$; again the flow of $g$ to $g^{\ast}$ is a power-law
in $\mu$, while that of $b$ and $\lambda$ is logarithmic:
\begin{equation}
\lambda (\mu) \sim - (\log(1/\mu))^{1/3}~~,~~
b(\mu) \sim (\log(1/\mu))^{-1/3}.
\end{equation}
The exponents at both the fixed line and the fixed point take the same values
as in (\ref{bosegasexponents}). However the flows above will lead to
logarithmic corrections which can be computed by standard methods. The
prefactors
of the these corrections, and also some amplitude ratios, will vary
continuously with the value of $b^{\ast}$ along the fixed line.

\section{Rotors with a non-zero mimimum angular momentum}
\label{monopole}
We now consider an extension of the rotor model $\hat{H}$ in which
each rotor has a `magnetic monopole' at the origin of $n$ space. Our
motivations for doing
this are: ({\em i}) It allows for quantized ferromagnetic states in which
the moment is a
half-integer times the number of rotor sites. Such phases will occur in
cases in which
each rotor corresponds to an odd number of Heisenberg spins in an
underlying
spin model. ({\em ii}) In $d=1$ (to be discussed in Sec~\ref{deq1})
such rotor models will lead, by the construction of Ref~\cite{shankar}, to
N\'{e}el phases
described by a sigma model with a topological theta term.

Following the method of Wu and Yang~\cite{wu}, we generalize the form
(\ref{lrep}) of $\hat{L}_\mu$ to
\begin{equation}
\hat{L}_{i\mu} = - \epsilon_{\mu\nu\lambda} n_{i\nu}
\left( \frac{\partial}{\partial n_\lambda} + q \varepsilon_i
A_{\lambda}(n_{i\mu}) \right) - q \varepsilon_i n_{i\mu}
\label{lrep2}
\end{equation}
where $q$ is chosen to have one of the values $1/2, 1, 3/2, 2, 5/2, \ldots$
and
$A_{\mu} (n)$ is the vector potential of a Dirac monopole at the origin
of $n$ space which satisfies $\epsilon_{\mu\nu\lambda}
\partial A_{\lambda}/ \partial n_{\nu} = n_{\mu}$ (we considered this
same
function in Sec~\ref{BtoD} for different reasons). We will work
exclusively on bipartite lattices and choose $\varepsilon_i = 1$
($\varepsilon_i = -1$) on the first (second) sublattice: we will comment
below
on the reason for this choice. It can now be verified that (\ref{lrep2})
continues to satisfy the commutation relations (\ref{commute}) for all
$q$.
However the Hilbert space on each site is restricted to states $|\ell ,
m\rangle$ satisfying $\hat{L}_\mu^2 |q,\ell , m \rangle = \ell (\ell + 1)
|q,\ell, m\rangle$ with $\ell = q , q+1, q+2, \ldots$ and $m = -\ell,
-\ell +1, \ldots, \ell$~\cite{wu}. Notice that there is a minimum value,
$q$, to the allowed angular momentum.

Another consequence of a non-zero $q$ is that it is no longer possible to
have ${\cal P}$ as a symmetry of a rotor Hamiltonian. There is a
part of $\hat{L}_\mu$ in (\ref{lrep2}) which is proportional to
$n_\mu$, and this constrains $\hat{L}_\mu$, $\hat{n}_\mu$ to have the
same
signature under discrete transformations: this rules out ${\cal P}$ as a
symmetry, even in models with $M=0$.

Our insertion of the $\varepsilon_i$ in (\ref{lrep2}) is also
related to the absence of the ${\cal P}$ symmetry for $q>0$.
Because $ \hat{L}_\mu$ and $\hat{n}_{\mu}$ have the same signature
under all
allowed symmetries for $q>0$, their expectation values turn
out to be proportional to each other on a given site, with the sign of the
proportionality
constant determined by
$\varepsilon_i$ (we will see this
explicitly
below). In a $q>0$ model with $\varepsilon_i = 1$ on every site,
the spatial average of
$\langle \hat{n}_\mu \rangle$ is proportional to that of
$\langle \hat{L}_\mu \rangle$
and therefore to the net ferromagnetic moment; as a result, such models
 turn out to have only
quantized ferromagnetic phases. Staggering of $\varepsilon_i$ is a way
of inducing a N\'{e}el-like order parameter with no net moment; then a
spatially uniform value of $\langle \hat{n}_{\mu} \rangle$ represents a
staggered mean value of the angular momentum $\langle \hat{L}_\mu
\rangle$, as expected from a N\'{e}el order parameter.
The staggering of $\varepsilon_i$ was also present
in the $d=1$
analysis of Ref~\cite{shankar}, and was crucial there in generating the
topological
theta term.

\begin{figure}
\epsfxsize=5.5in
\centerline{\epsffile{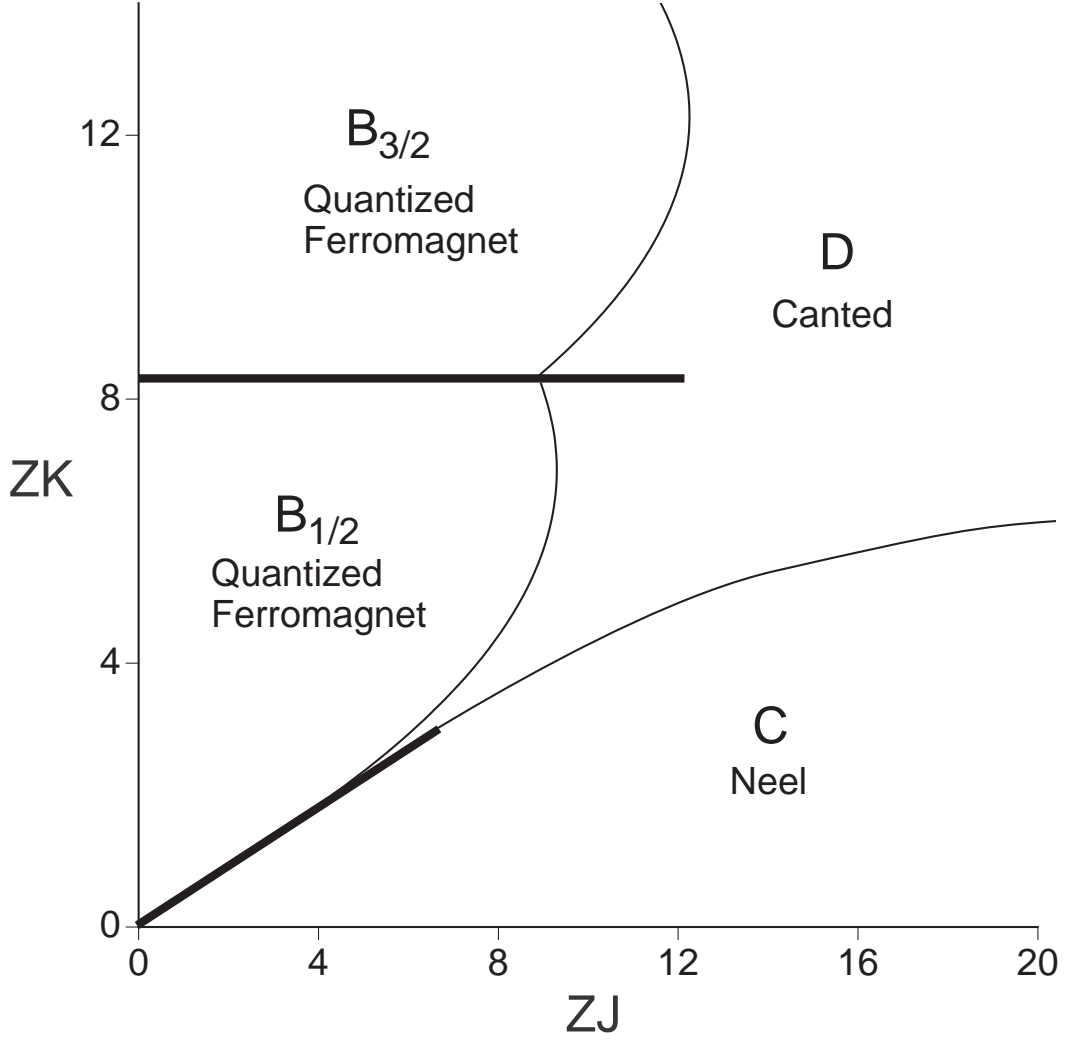}}
\vspace{0.5in}
\caption{Mean-field phase diagram of $\hat{H}$ with $\hat{L}_\mu$ given
by
(\protect\ref{lrep2}), monopole charge $q=1/2$, $M=0$ and other
parameters
and conventions are as in Fig~\protect\ref{phasediag1}.
The phases have the same physical properties as those of the
corresponding
phases in Fig~\protect\ref{phasediag1}. The moment, $\ell$, of the
quantized
ferromagnet phases $\rm{B}_\ell$ is now half-integral, and such phases
exist for all half-integers $\ell$ at larger $K$. The model is on a bi-
partite
lattice, and the expectation values (up to a global $O(3)$ rotation)
in the various phases  on the two
sublattices, 1,2, are:
($\rm{B}_{\ell}$) $\langle\hat{L}_{1z} \rangle = \langle\hat{L}_{2z}
\rangle = \ell $,
$\langle \hat{n}_{1z} \rangle = -\langle \hat{n}_{2z} \rangle\neq 0$,
$\langle\hat{L}_{1x,1y,2x,2y} \rangle =0$,
$\langle \hat{n}_{1x,1y,2x,2y} \rangle = 0$;
(C) $\langle\hat{L}_{1x} \rangle = -\langle\hat{L}_{2x}
\rangle \neq 0 $,
$\langle \hat{n}_{1x} \rangle = \langle \hat{n}_{2x} \rangle\neq 0$,
$\langle\hat{L}_{1y,1z,2y,2z} \rangle =0$,
$\langle \hat{n}_{1y,1z,2y,2z} \rangle = 0$;
(D) $\langle\hat{L}_{1z} \rangle = \langle\hat{L}_{2z}
\rangle \neq 0 $,
$\langle \hat{n}_{1z} \rangle = - \langle \hat{n}_{2z} \rangle\neq 0$,
$\langle\hat{L}_{1x} \rangle = - \langle\hat{L}_{2x}
\rangle \neq 0 $,
$\langle \hat{n}_{1x} \rangle =  \langle \hat{n}_{2x} \rangle\neq
0$, $\langle\hat{L}_{1y,2y} \rangle =0$,
$\langle \hat{n}_{1y,2y} \rangle = 0$;
}
\label{phasediag2}
\end{figure}

We present in Fig~\ref{phasediag2} the mean field phase diagram of the
Hamiltonian
$\hat{H}$ (Eqn (\ref{hrotor})) on a bipartite lattice with $M=0$,
$\hat{L}_\mu$ given by (\ref{lrep2}), and $q=1/2$. Because of absence of
the ${\cal P}$ symmetry, a non-zero $M$ will not make a qualitative
difference.
Other values of the $q$ are also expected to have a similar phase diagram.
The
caption describes the non-zero components of $\langle \hat{L}_\mu
\rangle$
and $\langle \hat{n}_\mu \rangle$ in the various phases.
The expectation values
$\langle n_\mu \rangle$ have exactly the same form as the $\langle
\hat{L}_\mu \rangle$ in spin space, but the opposite signature in
sublattice
space ({\em e.g.} a staggered configuration of $\langle \hat{L}_x \rangle$
implies a uniform configuration of $\langle \hat{n}_x \rangle$ and vice
versa).

The phases in Fig~\ref{phasediag2} are closely related to those
in Fig~\ref{phasediag1}. As before we have the quantized ferromagnetic
phases B (albeit, now with half-integral moments), the N\'{e}el phase C,
and the canted phase D. The main difference is in the absence of a
quantum paramagnetic phase A: this is clearly due to the minimum allowed
value $q=1/2$ in the single rotor angular momentum.

It is now possible to undertake an analysis of the low energy properties of
the phases, and of the critical properties of the phase transitions, much
like that carried out in Secs~\ref{struct} and~\ref{qpt} for the $q=0$
case: such an analysis shows essentially no differences between  $q=0$
and $q>1$, at least for $d>1$. The excitations of the phases
B, C, and D in Fig~\ref{phasediag2} are the same as those
of the corresponding phases in Fig~\ref{phasediag1}, as are the
universality classes of the continuous phase transitions between C and D,
and between B and D. One small, but important, point has to be kept in
mind in
this regard. The $q>0$ models are not invariant under the symmetry
${\cal P}$ even when $M=0$, and so the restrictions that the ${\cal P}$
symmetry implies for the $q=0$ analysis of Secs~\ref{struct}
and~\ref{qpt} must not be  imposed now.

In $d=1$, there are significant differences between the $q=0$ and $q=1/2$
cases, and these will be discussed in the next section.

\section{Quantum rotors in one dimension}
\label{deq1}
The general topology of the mean-field phase diagrams
Figs~\ref{phasediag1}
and~\ref{phasediag2} is expected to hold for all $d>1$. As we have seen
in Sec~\ref{qpt}, for $d<2$ fluctuations do modify the critical properties
of the continuous transitions, but these modifications are computable in a
systematic expansion in $2-d$. In $d=1$, fluctuations modify not only the
transitions, but also the stability of the phases: as a result, we
expect significant changes in the topology of the phase diagram itself.
Further, we also expect a sensitive dependence to the value of the
monopole charge $q$. In the following, we present a mixture of results,
educated surmises and speculation
on the nature of the phases and phase transitions in $d=1$
for different values of $q$.

\subsection{$q=0$}
\begin{figure}
\epsfxsize=5.5in
\centerline{\epsffile{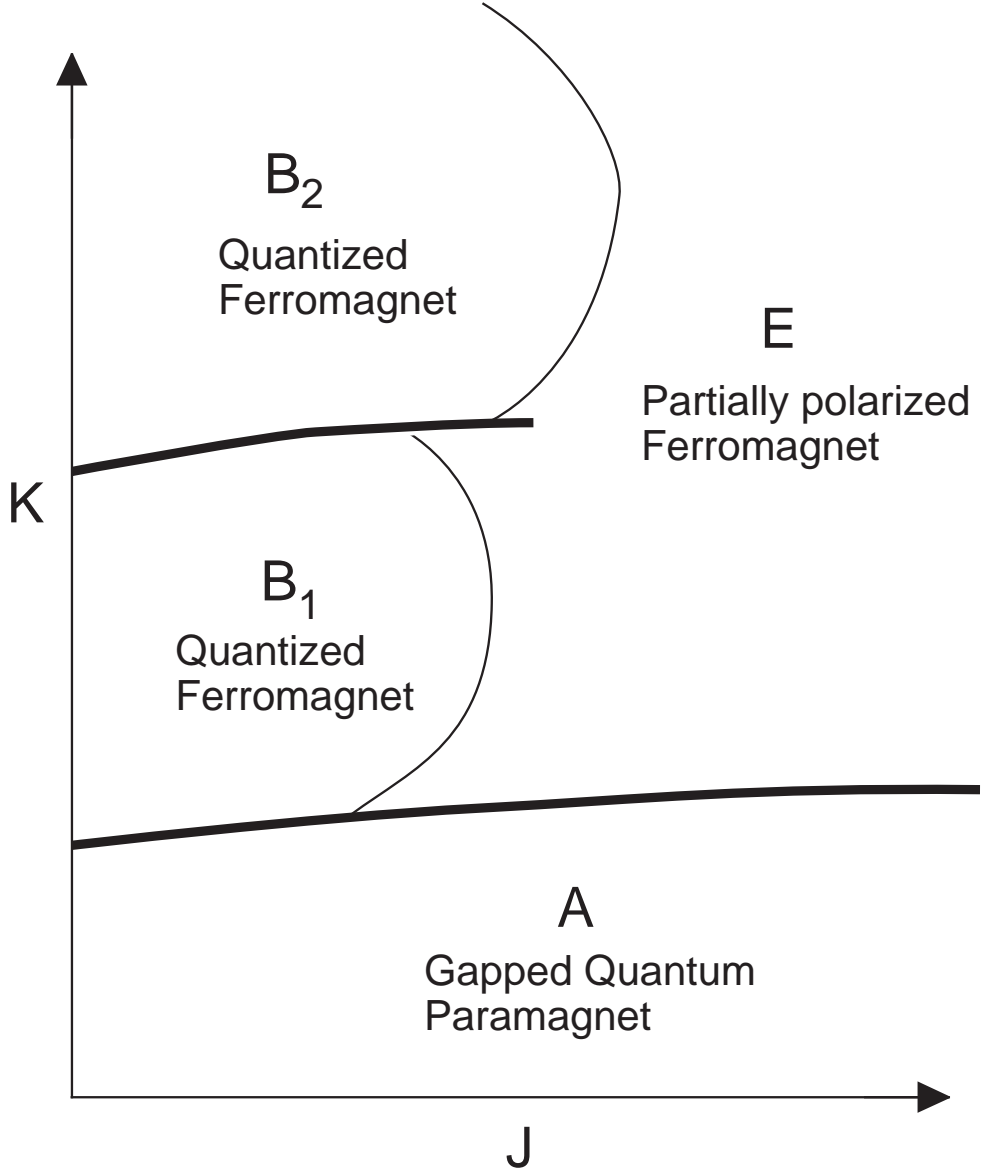}}
\vspace{0.5in}
\caption{Expected phase diagram of $\cal{H}$ in $d=1$ with monopole
charge $q=0$. The phases A and $\rm{B}_{\ell}$ are as in
Fig~\protect\ref{phasediag1}. The $d>1$ phase D becomes phase
E in $d=1$; the latter phase has $\langle \hat{n}_\mu \rangle =0$,
$\langle \hat{L}_z \rangle \neq 0$, $\langle \hat{L}_{x,y} \rangle = 0$.
The
magnetic moment is not quantized but varies continuously. The
$\hat{n}_{x,y}$
fields have algebraic correlations in space.
}
\label{q0}
\end{figure}

The expected phase diagram is shown in Fig~\ref{q0}. We discuss some of
the
important features in turn:\\
({\em i}) There is no phase with N\'{e}el order (the analog of phase C for
$d>1$);
it has been pre-empted mostly by the quantum paramagnet A. Fluctuations
in the incipient N\'{e}el phase would be described by a $O(3)$ non-linear
sigma
model in $1+1$ dimensions (without a topological term), which is known
not to have a phase with
long-range order. \\
({\em ii}) The quantized ferromagnetic phases, B, are stable even in $d=1$.
They have the usual $\omega \sim k^2$ spectrum of spin-wave
excitations.\\
({\em iii}) The canted phase D has been replaced by a novel new phase
E - the partially polarized ferromagnet. The phase E has true long-range
order in the ferromagnetic order parameter $\langle \hat{L}_z \rangle$;
however the value of $\langle \hat{L}_z \rangle$ is not quantized and
varies
continuously. The canted phase D, in $d>1$, also had long-range order
in $\hat{n}$ in the $x-y$ plane; in contrast, in phase E ($d=1$) this long-
range
order has been replaced by a quasi-long-range XY order {\em i.e.}
correlations
of $\hat{n}$ decay algebraically in the $x-y$ plane. The only
true broken symmetry is that associated with $\langle \hat{L}_z \rangle$,
and the system is invariant with respect to rotations about the $z$ axis
(compare with phase D in which the $O(3)$ symmetry was
completely broken and there was no invariant axis).
Note that though there is no LRO in the $x-y$ plane in phase E,
the Goldstone mode of phase D, associated with it's $x-y$ LRO
survives i.e. phase E has not only a gapless mode
$\omega \sim k^2$ due to ferromagnetic long-range order, but also
a gapless mode $\omega \sim k$ due to quasi LRO in the $\hat{n}$
field. All of these results follow from a straightforward analysis of the
actions (\ref{sigmaBtoD}) and (\ref{lsigma2}) in $d=1$.\\
({\em iv}) A few remarks about the universality class of the phase
transition
between phases B and E. The theory (\ref{lsigma2}) should continue to
apply.
The expansion in $\epsilon=2-d$ (Section~\ref{secrgBtoD}) should be valid
all the way down to
$d=1$~\cite{sss}, although it may not be quantitatively accurate.
The theory with ${\cal P}$ symmetry has critical properties identical to a dilute
Bose gas, and was discussed in Ref~\cite{sss} and
solved there by a fermionization trick. The theory without ${\cal P}$ symmetry,
has non-analytic terms in the action, and is probably not amenable to a simple
solution by fermionization.\\  
({\em v}) The phase transition between phase A
and phase E is always expected
to be first-order. The quantum paramagnetic phase A has no gapless
excitations
and a vanishing spin susceptibility. It is then difficult to conceive
of a mechanism which could lead to a continuous condensation of the field
$h_\mu$ in an action like (\ref{landau}) and (\ref{lsigma1}) (or
(\ref{landauM}) for $M \neq
0$): integrating out the $N_\mu$ fluctuations does not yield a negative
contribution to the
``mass'', $r_1$, of the $h_\mu$ field, as the spin susceptibility is zero.

\subsection{Integers $q > 0$}
\begin{figure}
\epsfxsize=5.5in
\centerline{\epsffile{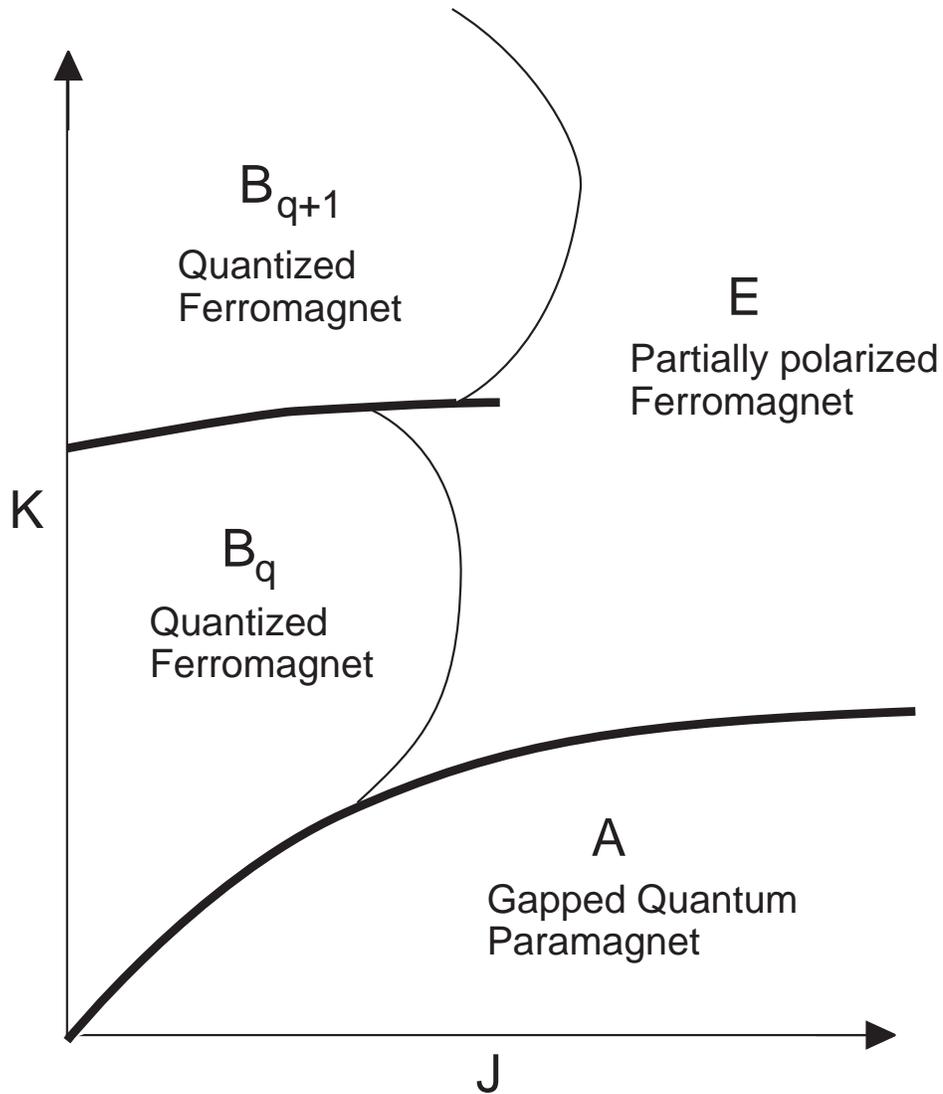}}
\vspace{0.5in}
\caption{Expected phase diagram of $\cal{H}$ in $d=1$ with monopole
charge an integer $q>0$. Phases are similar to those with the same labels
in Fig~\protect\ref{phasediag2} and~\protect\ref{q0}.}
\label{q1}
\end{figure}

The phase diagram is shown in Fig~\ref{q1}. All of the phases are
identical to
those discussed above for $q=0$. The only difference is in the behavior of
the first-order line surrounding phase A: it bends down towards the origin
of the $J$-$K$ plane, as the minimum possible value of the single site
angular momentum always forces in quantized ferromagnetic phases for
$J$ small
enough.

\subsection{Half-integers $q>0$}
\begin{figure}
\epsfxsize=5.5in
\centerline{\epsffile{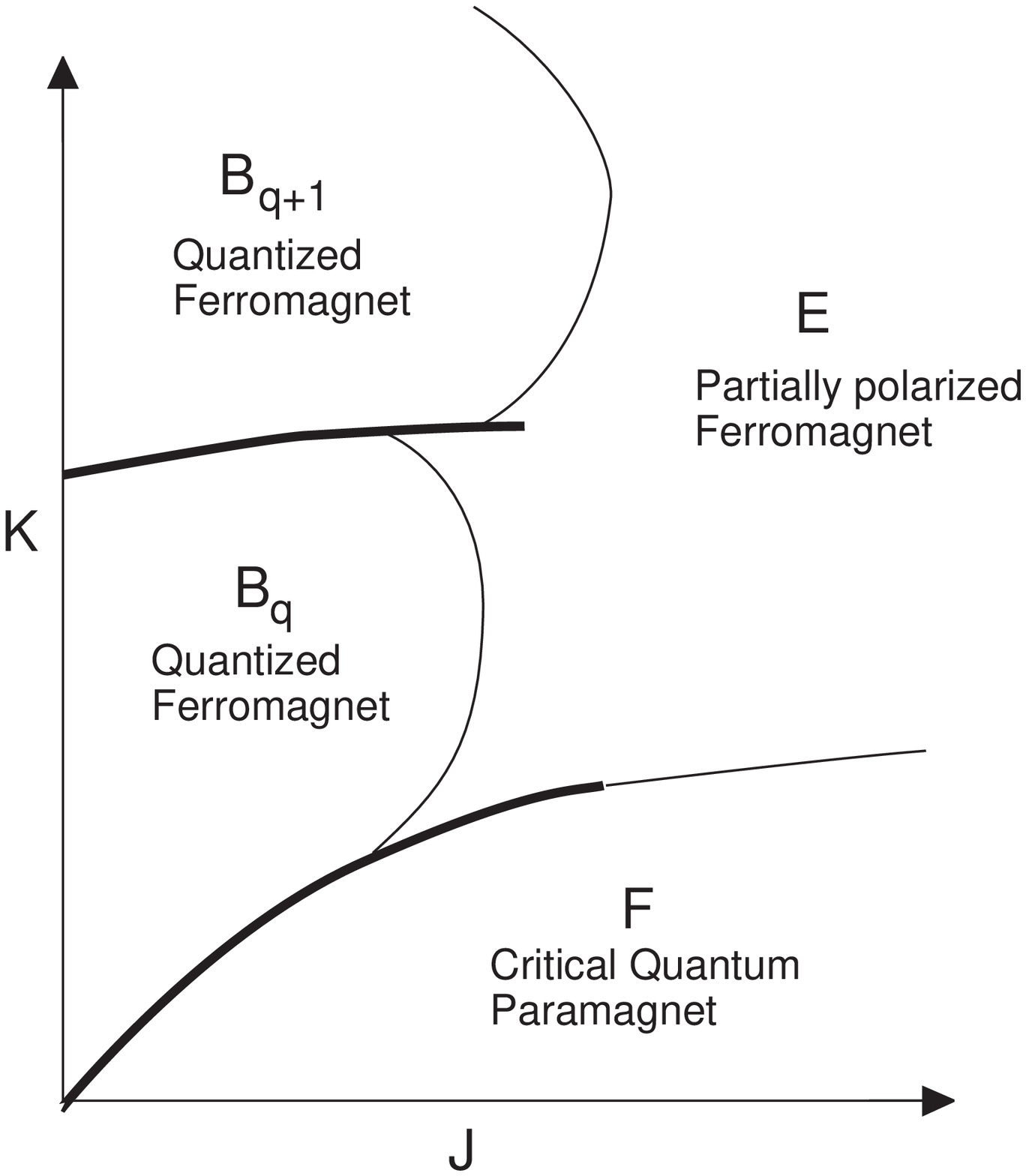}}
\vspace{0.5in}
\caption{Expected phase diagram of $\cal{H}$ in $d=1$ with monopole
charge a half-integer $q>0$. Phases are similar to those with the same
labels
in Fig~\protect\ref{phasediag2} and~\protect\ref{q0}.
Phase F has no broken symmetry and algebraic correlations of all spin
operators. Its low energy properties are described by the $1+1$
dimensional
$O(3)$ non-linear sigma model with a topological term with co-efficient
$\theta = \pi$. }
\label{qh}
\end{figure}

The phase diagram is shown in Fig~\ref{qh}. The topology is now similar
to the positive integer case, as are the phases B and E, and the
transitions between them. The primary difference is that the quantum
paramagnet A has been replaced by a critical phase F, which has no
broken symmetries and power-law decay of all observables.
Fluctuations in phase F are described by the $1+1$ dimensional $O(3)$
non-linear
sigma model with a topological term with coefficient $\theta=\pi$: this
mapping follows from the analysis of Ref~\cite{shankar}.

Finally, we make a few remarks on the transition between the critical
phase
F and the partially polarized ferromagnet E. As F has gapless excitations,
this transition can be continuous. A (strongly-coupled) field theory for
this transition is given by the action ${\cal L}_1$ in (\ref{landauM}),
supplemented by the constraints (\ref{constraintsM}) and a topological
term at $\theta =\pi$ in the unit-vector $N_\mu$ field.
An alternative, Hamiltonian point-of-view on the same transition is the
following. It is known~\cite{affleckhaldane} that the critical $1+1$
dimensional $O(3)$
non-linear sigma model is equivalent to the $SU(2)$ Wess-Zumino-Witten
model~\cite{wzw}
at level $k=1$. The Hamiltonian of this model is given by
\begin{equation}
H_{WZW} = \alpha (\vec{J}_L^2 + \vec{J}_R^2)
\end{equation}
where $\vec{J}_{L,R}$ are the $SU(2)$ currents obeying a Kac-Moody
algebra.
We now want to induce a ferromagnetic moment into the ground state of
this theory. The action (\ref{landauM}) does this by coupling in a
fluctuating magnetic field $h_\mu$. Such a field would couple here
to the magnetization $\vec{J}_L + \vec{J}_R$: integrating out
$h_\mu$ would then induce a coupling $- (\vec{J}_L + \vec{J}_R)^2$ which
gives
us the Hamiltonian~\cite{foot}
\begin{equation}
H_{F\rightarrow E} = \alpha' (\vec{J}_L^2 + \vec{J}_R^2) - \lambda
\vec{J}_L \cdot \vec{J}_R .
\label{wzwa}
\end{equation}
The model $H_{F\rightarrow E}$ can also be used to describe the onset of
ferromagnetism in an itinerant Luttinger liquid. 
It has in fact been examined earlier by
Affleck~\cite{affleck}, where he obtained it as an effective model for the
spin degrees
of freedom in a $d=1$ Hubbard model.
Affleck examined the RG flows of $\lambda$ for small $\lambda$ and
obtained
\begin{equation}
\frac{d \lambda}{d \ell} = - \lambda^2
\label{lflow}
\end{equation}.
In the mapping from the repulsive Hubbard model, and also from our rotor
model, the
initial sign of
$\lambda$ is positive. From (\ref{lflow}), Affleck concluded that
$\lambda$ is irrelevant for all positive $\lambda$, and that all such
systems flow into the $\lambda = 0$ fixed point. We believe this
conclusion is
incorrect. It is clear from our arguments that for $\lambda$ large
enough, $H_{F\rightarrow E}$ should undergo a phase transition to
ferromagnetic
ground state. This suggests that there is a critical value
of $\lambda = \lambda_c$ (with $\lambda_c >0$ and of order unity) such
that
only systems with $\lambda < \lambda_c$ flow into the $\lambda = 0$
fixed
point. Systems with $\lambda > \lambda_c$ are in the ferromagnetic
phase E.
The nature of the critical point at $\lambda = \lambda_c$, which controls
the transition from phase F to E, is unknown: determining its structure
remains an important open problem.

\section{Conclusions}
\label{conc}
This paper has introduced and analyzed the simplest model with
Heisenberg symmetry which
exhibits zero temperature phase transitions, and whose phases contain a
net average magnetic moment. The model contained only bosonic quantum
rotor degrees of freedom and offers the simplest realization of a  quantum
transition with
an order parameter which is also a non-abelian, conserved charge. The
analysis focussed
primarily on the Hamiltonian $\hat{H}$ in (\ref{hrotor}), although
variations were also
considered. The results are summarized in the phase diagrams in
Figs~\ref{phasediag1}-\ref{qh}.

Some important properties of these phase diagrams deserve reiteration.
Notice that for $d>1$, there is no phase which is simply a non-quantized
ferromagnet, with no other broken symmetry. Phase D has a non-quantized
ferromagnetic moment, but it has an additional long-range order in the $n$
field in a plane perpendicular to the ferromagnetic moment. We believe this is
a generic feature of insulating spin systems: ferromagnetic ground states
either have an integral or half-integral magnetic moment, or have an additional
broken symmetry. Only in $d=1$ does a phase like E appear: it has a
non-quantized ferromagnetic moment, and is invariant under rotations about the
ferromagnetic axis. However, even in $d=1$ there is a remnant of the broken
symmetry in the $n$ field perpendicular to the moment: correlators of $n$ have
a power-law decay in space, and the linearly-dispersing gapless spin-wave mode
is still present (in addition to the usual quadratically dispersing
ferromagnetic mode). It is also interesting to note that metallic, Fermi liquids
of course have no trouble forming non-quantized ferromagnets; this is in
keeping with popular wisdom that Fermi liquids are ``effectively'' one
dimensional. 

A second interesting property of the phase diagram was pointed out in
Section~\ref{intro}: continuous zero
temperature transitions in
which there is an onset in the ferromagnetic moment
only occur from phases
which have gapless excitations. Thus there is such an onset from phase C
(which has gapless spin waves) to phase D, but no continuous transition between
phases A and D.

 We also examined the
critical properties of the second order phase transitions in the model. In several
cases, the critical theories turned out to be  variations on the theme of a simpler
quantum phase transition: the onset of density in a Bose gas with repulsive
interactions as its chemical potential is moved through zero. This quantum transition
had been studied earlier~\cite{fwgf,sss}. Because of its central importance, we
obtained some additional results on its
$T=0$ universal properties in Appendix~\ref{appcalc}.

In the remainder of this section,
we remark on issues related to those considered in this paper, but which we have not
directly analyzed here.

\subsubsection{Itinerant ferromagnets}
We discuss implications of our results for quantum phase transitions in
ferromagnetic Fermi liquids. As noted in Section~\ref{intro}, there are two quantum
transitions in this system, and we will discuss them separately:
\newline
({\em i}) Consider first the transition from a fully polarized Fermi liquid to a
partially polarized Fermi liquid. This is rather like the transition from 
phase B to phase D in $d>1$, and the transition from phase B to phase E in $d=1$.
Let us suppose that all electrons in the fully polarized state are polarized in
the `up' direction. Then an order parameter for the transition is simply the
density of `down' spin electrons. Along the lines of the analysis in the rotor
model, we can derive an effective action for the down spin electrons simply by
integrating out the up electrons. It turns out that the up electrons only mediate
irrelevant interactions between the down electrons: the effective action for the
down electrons is simply that of a dilute gas of {\em free } spinless
fermions~\cite{conserve,china}. A possible four-point interaction like 
the $|\psi|^4$ term for the Bose gas is prohibited by Fermi statistics;
further a singular term, like that appearing in the models without ${\cal P}$
symmetry in Section~\ref{secrgBtoD}b, appears to be prohibited here because the
large Fermi momentum of the up electrons inhibits strong mixing with down electrons
via emission of ferromagnetic spin waves.
The properties of the free spinless fermion
model are of course trivial, but it is quite useful to re-interpret them in the
language of a quantum phase transition~\cite{china}. It is also interesting to note
here that the  critical theory of free spinless fermions in $d=1$ is identical to
that of dilute interacting bosons for the B to E transition in $d=1$~\cite{sss}. This
is in keeping with our assertion that in $d=1$, the transitions in itinerant fermion
systems are in the same universality class as those in certain rotor models. 
\newline
({\em ii}) Consider now the onset of ferromagnetism in an unpolarized Fermi liquid.
A theory for this transition for $d>1$ was proposed by Hertz~\cite{hertz}.
This transition is similar to the transition between phases C and D studied in
Section~\ref{CtoD}. In the end, our analysis used a method very close in spirit to
that use by Hertz: simply integrate out all gapless modes not directly related to
the order parameter. We have provided here some independent justification for such
an approach in the rotor model, and our results  provide support for the
correctness of Hertz's analysis for $d>1$. 
Precisely in $d=1$, we have no theory for the transition between phases
F and E, but we have argued that it should be in the same universality class 
as that of the onset of ferromagnetism in a Luttinger liquid of
itinerant electrons. The critical theory of this transition is the main remaining
open problem in the theory of phase transitions in quantum ferromagnets.

\subsubsection{Finite temperature}
Essentially all the analysis of this paper was at
$T=0$, and it would be interesting to extend it to finite $T$. The finite
$T$ properties
of the quantum paramagnetic phase A, the N\'{e}el phase C, and the
transition between
them, have already been studied in some detail~\cite{CHN,CSY}. More
recently, a
field-theoretic analysis of the finite $T$ properties above a quantized
ferromagnetic
state B has appeared~\cite{ferroqhe}. It remains, therefore, to study the
canted phase D
and its phase transitions at finite $T$. We have shown in this paper that
the transition
between the B and D phase is similar (although not identical) to that in a
dilute Bose
gas~\cite{fwgf,sss}---the finite $T$ analysis should therefore be related
to that in
Ref~\cite{sss}. The finite $T$ properties of the phase D itself should also
be quite
interesting, especially in $d=2$ where the long-range order disappears at
any finite
$T$. This is also the case of direct relevance to the quantum Hall
effect experiments of Ref~\cite{barrett}.
The phase D could be modeled {\em e.g.} by the action (\ref{sigmaBtoD}),
supplemented
by the additional constraint $\hat{n}_{\mu}^2 = 1$, which can be imposed
once we are
well away from phase B.

\subsubsection{Effects of randomness}
\label{randomness}
At $T=0$, but in the presence of randomness, one might expect that between the
quantized ferromagnetic (B) and canted phases (D), there occurs a ``canted glass''
phase with a non-quantized magnetic moment, short-range mean N\'{e}el correlations,
but a diverging mean N\'{e}el susceptibility; this can be seen by arguments
analogous to those in Ref~\cite{fwgf}. Further, general arguments~\cite{berker}
imply that for arbitrarily weak randomness in $d \leq 2$ and for sufficiently strong
randomness in $d>2$, all the first order lines in our phase diagram are replaced by
second-order transitions (implying, for instance, the absence of direct transitions
between the quantized ferromagnetic phases).

\acknowledgements
We thank A. Chubukov and N. Read for useful discussions.
This research was supported by NSF Grant No. DMR-92-24290.

\appendix

\section{Mapping between Heisenberg spin and rotor models}
\label{twolayer}

Consider a `double layer' model of spin $S$ Heisenberg spins
$\hat{S}_{ia\mu}$,
$\hat{S}_{ib\mu}$ on the sites $i$ of a $d$ dimensional lattice; $a,b$ are
two `layer'
indices and $\mu=1,2,3$ are vector components. We study the Hamiltonian
\begin{equation}
\hat{H}_{dl} = G \sum_{i,\mu} \hat{S}_{ia\mu} \cdot \hat{S}_{ib\mu}
- \sum_{<ij>,\mu} \left( J_a \hat{S}_{ia\mu} \cdot \hat{S}_{ja\mu}
+ J_b \hat{S}_{ib\mu} \cdot \hat{S}_{jb\mu}
+ J_{ab} \left( \hat{S}_{ia\mu} \cdot \hat{S}_{jb\mu} + 
\hat{S}_{ja\mu} \cdot \hat{S}_{ib\mu} \right)\right),
\end{equation}
where $<ij>$ is a sum over nearest-neighbor pairs.
We will consider the case where $G$ is antiferromagnetic ($G > 0$).
Then, neglecting the inter-site terms, each site has a tower of states
with total angular
momentum $\ell = 0, 1, 2, \ldots 2S$. This tower is very similar to that of
a single
quantum rotor, the main difference being that the latter does not have an
upper bound on
its allowed angular momentum. For the low energy properties of interest
in this paper,
the upper bound is not expected to be important. Further, by constructing
an on-site
Hamiltonian which is a polynomial in $\hat{L}_{\mu}^2$  ($\hat{L}_{\mu}$
is the rotor
angular momentum), it is possible to mimic the actual eigenenergies of the
tower of
states in the Heisenberg system. 

Now consider the inter-site terms in $H_{dl}$. For the on-site tower, the
matrix
elements of the rotor operators $\hat{L}_{\mu}$ and $\hat{n}_{\mu}$ are
similar to those
of $\hat{S}_{a\mu} + \hat{S}_{b\mu}$ and $f(\hat{S}_{a\mu} -
\hat{S}_{b\mu})$
respectively ($f$ is some constant); this correspondence becomes exact in
the
semiclassical theory. So we perform the replacement $\hat{S}_{a\mu}
= (\hat{L}_\mu + f \hat{n}_\mu)/2$, $\hat{S}_{b\mu}
= (\hat{L}_\mu - f \hat{n}_\mu)/2$ in $\hat{H}_{dl}$. This yields
precisely the inter-site
terms in (\ref{hrotor}) with
\begin{equation}
K = \frac{J_a + J_b + 2J_{ab}}{4}~~;~~
J = f^2\frac{J_a + J_b - 2J_{ab}}{4}~~;~~
M = f\frac{J_a - J_b}{4}.
\end{equation}
Note that in systems with a layer interchange symmetry ($J_a = J_b$), the
coupling $M$
vanishes. Thus this interchange symmetry is equivalent to ${\cal P}$.

\section{Mean Field Theory}
\label{mfta}
\subsection{Rotors with $q = 0$}
The Hamiltonian
$\hat{H}_{mf}$ of Section~\ref{mft} was diagonalized
by determining its matrix elements in
the basis of spherical harmonic states: $| \ell, m \rangle$ with
$\ell, m$ integers satisfying $-\ell \leq m \leq \ell$ and $\ell \geq 0$.
The matrix elements of the operators in this basis can be expressed
in terms of Clebsch-Gordon co-efficients; the non-zero matrix elements
are:
\begin{eqnarray}
\langle \ell, m | \hat{L}_z | \ell m \rangle &=&
 m \nonumber \\
\langle \ell, m+1 | \hat{L}_{+} | \ell m \rangle &=&  ((\ell + m
+ 1)(\ell - m))^{1/2} \nonumber \\
\langle \ell, m | \hat{n}_{z} | \ell+1, m \rangle &=&
 \left( \frac{(\ell + m+1)( \ell -m +1)}{(2 \ell +1) (2 \ell + 3)}
\right)^{1/2} \nonumber \\
\langle \ell, m+1 | \hat{n}_{+} | \ell+1, m \rangle &=&
\left( \frac{(\ell - m)( \ell -m +1)}{(2 \ell +1) (2
\ell + 3)}
\right)^{1/2}\nonumber \\
\langle \ell, m -1| \hat{n}_{-} | \ell+1, m \rangle &=&
- \left( \frac{(\ell + m)( \ell + m +1)}{(2 \ell +1) (2 \ell + 3)}
\right)^{1/2} ,
\label{matelem}
\end{eqnarray}
and their complex conjugates.
Here, as usual, $\hat{L}_{+} = \hat{L}_x + i\hat{L}_y$ and similarly for
$\hat{n}_{+}$. All matrix elements not obtainable by complex conjugation
of the above are zero.
notice that $\hat{n}_\mu$ only has non-zero matrix elements between
states
with angular momenta $\ell$ and $\ell \pm 1$. In particular, matrix
elements of $\hat{n}_\mu$ between states with the same value of $\ell$
vanish; this happens because $\hat{n}_\mu$ is odd under ${\cal P}$,
and the $|\ell, m\rangle$ states have definite ${\cal P}$-parity.

The single-site Hilbert space has an infinite number of states, but
in practice it was found that good accuracy was obtained by truncating the
states above a maximum value of $\ell \approx 15$.
The results of the numerical calculation were discussed in
Section~\ref{intro}
and in Fig~\ref{phasediag1}.

\subsection{Rotors with $q = 1/2$}
The mean-field analysis proceeds in a manner similar to Sec~\ref{mft}.
We now have to introduce two distinct mean field Hamiltonians
$\hat{H}_{1mf}$, $\hat{H}_{2mf}$ for the two sublattices, each with
their own effective fields $N_{1\mu}$, $h_{1\mu}$ and $N_{2\mu}$,
$h_{2\mu}$.
The expectation value of $\hat{H}$ in the ground state of the
mean-field Hamiltonians is then minimized with respect to variations
in these four effective fields. The numerical diagonalization of the mean-
field
Hamiltonians requires the matrix elements of the operators in the
states of rotor with a monopole $q$. We determined these for $q=1/2$
from
Ref~\cite{wu} and applications of the Wigner-Eckart theorem.
The matrix elements of the $\hat{L}_{\mu}$ are still given by those in
(\ref{matelem}) ({\em i.e.} they are independent of $q$) while those of the
$\hat{n}_\mu$ are
\begin{eqnarray}
\langle 1/2,\ell, m' | \hat{n}_{i\mu} | 1/2,\ell m \rangle &=&
- \frac{\varepsilon_i}{\ell (2 \ell + 2)} \langle 1/2,\ell, m' | \hat{L}_\mu |
1/2,\ell m
\rangle
\nonumber \\
\langle 1/2,\ell, m | \hat{n}_{iz} | 1/2,\ell+1, m \rangle &=&
 \frac{\varepsilon_i}{2\ell + 2}
\left( \frac{(\ell + m+1)( \ell -m +1)(2\ell +1)}{2}
\right)^{1/2} \nonumber \\
\langle 1/2,\ell, m +1 | \hat{n}_{i+} | 1/2,\ell+1, m \rangle &=&
 \frac{\varepsilon_i}{2\ell + 2}
\left( \frac{(\ell - m)( \ell -m +1)(2\ell +1)}{2}
\right)^{1/2} \nonumber \\
\langle 1/2,\ell, m -1 | \hat{n}_{i-} |1/2, \ell+1, m \rangle &=& -
 \frac{\varepsilon_i}{2\ell + 2}
\left( \frac{(\ell + m)( \ell + m +1)(2\ell +1)}{2}
\right)^{1/2} .
\end{eqnarray}
All matrix elements not obtainable from the above by complex conjugation
are zero.
Notice that, unlike (\ref{matelem}), $\hat{n}_\mu$ now has
non-zero matrix elements between states with the same value of $\ell$,
and these matrix elements are proportional to those of $\hat{L}_\mu$;
this happens because the $|q>0,\ell , m\rangle$ states no longer have
definite ${\cal P}$ parity.
The results of this calculation were discussed in
Section~\ref{monopole}.

\section{Computations for N\'{e}el (C) to Canted (D) Transition}
\label{apprg1}

We will discuss the case without ${\cal P}$ symmetry, described by the action
(\ref{actionmneq0}) in Section~\ref{secrgmneq0}. The results with ${\cal P}$
symmetry in Section~\ref{secrgmeq0} follow as the special case $\lambda_0 = 0$.

At the critical point, the propagator of (\ref{actionmneq0}) is (dropping the
$0$ subscripts on the couplings)
\begin{eqnarray}
G(x,\tau) &=& \int \frac{d^d k}{(2 \pi)^d} \frac{d \omega}{2 \pi}~
\frac{e^{i (\vec{k} \cdot \vec{x} - \omega \tau)}}{
k^2 + b \omega^2 / k^2 - i \lambda b^{1/2} \omega} \nonumber \\
&=& \frac{1}{(b(\lambda^2 + 4))^{1/2}}
\frac{1}{(4 \pi |\tau|)^{d/2}} \left( \frac{e^{-x^2 /4 \lambda_+ \tau}}{
\lambda_+^{d/2}} \theta (\tau) + 
\frac{e^{-x^2 /4 \lambda_- |\tau|}}{
\lambda_-^{d/2}} \theta (-\tau) \right)
\end{eqnarray}
where
\begin{equation}
\lambda_{\pm} = \frac{ (\lambda^2 + 4)^{1/2} \mp \lambda}{2 b^{1/2}}.
\end{equation}
The order $u^2$ contribution to the self energy is therefore
\begin{equation}
\Sigma (k, \omega) = 2 u^2 \int d^d x d \tau G^2 (x, \tau) G(-x, -\tau)
e^{-i (\vec{k} \cdot \vec{x} - \omega \tau)}
\end{equation}
After a lengthy, but straightforward, evaluations of the above integral
we find, to leading order in $\epsilon = 2-d$:
\begin{equation}
\Sigma (k, \omega ) - \Sigma (0,0) = 
- \frac{u^2}{8 \pi^2 b \epsilon} \frac{(2 \lambda^2 + 9)(-i \lambda b^{1/2}
\omega + k^2) - \lambda^2 k^2}{(\lambda^2 + 4)(2 \lambda^2 + 9)^2} + \ldots
\end{equation}
The renormalization constants $Z$, $Z_b$, and $Z_{\lambda}$ in (\ref{Zvalues})
follow immediately from the above result.

Similarly, the four-point vertex is
\begin{eqnarray}
&& u - u^2 \int d^d x d \tau G^2 (x, \tau) - 
4 u^2 \int d^d x d \tau G (x, \tau) G(-x, -\tau) \nonumber \\
&&~~~~~~~~= u - \frac{u^2}{4 \pi b^{1/2} \epsilon} \frac{\lambda^2 +
20}{(\lambda^2 + 4)^{3/2}} + \ldots,
\label{calcz4}
\end{eqnarray}
which leads to the result for $Z_4$ in (\ref{Zvalues}).

The computation of $Z_2$ is very similar, and details are omitted.

\section{Dilute Bose gas below two dimensions}
\label{appcalc}

In this Appendix we will study the Bose gas described by the action
\begin{equation}
{\cal S} = \int d^d x d\tau \left( \psi^{\ast} \frac{\partial \psi}{\partial
\tau} + | \nabla \psi |^2 + r |\psi |^2 + \frac{u}{2} |\psi|^4 \right).
\end{equation}
This action undergoes a $T=0$ quantum phase transition, at $r=0$, which plays an
important role in the models considered in the main part of this paper.
In $d<2$, this quantum transition obeys a {\em no scale-factor
universality\/}~\cite{sss} which we shall study here in greater detail.
One consequence of this enhanced universality in $d<2$ is that the
zero temperature density,
$n = \langle |\psi |^2 \rangle$, of this Bose gas obeys 
\begin{equation}
n= {\cal C}_d
\theta(-r) |r|^{d/2},
\label{ncalc}
\end{equation}
with ${\cal C}_d$ a universal number~\cite{sss}.
Here we shall show how to compute ${\cal C}_d$ in an expansion in powers of $\epsilon
= 2 - d$. It is known that the leading term is ${\cal C}_d = 1/(4 \pi \epsilon)$,
and we shall explicitly determine the next term. We shall also compute the effective
potential of the Bose gas to order $\epsilon$.

We characterize the interactions in ${\cal S}$ by the bare dimensionless coupling
$g_0$ defined by
\begin{equation}
g_0 = \mu^{-\epsilon} S_d u
\label{ug0}
\end{equation}
where $\mu$ is a momentum scale, and recall that
$S_d = 2 \pi^{d/2} /((2 \pi)^2 \Gamma (d/2))$. We define a renormalized
dimensionless coupling $g$ in a similar manner by replacing $u$ with the value of
the exact two-particle scattering amplitude at zero external frequencies and  equal
incoming momenta $p$, with $p^2 = s^2 \mu^2$. The dimensionless number $s$ is
arbitrary, and no universal quantity should depend upon it; we shall keep track of
the $s$ dependence as a check on  the universality of our final results. The
two-particle scattering amplitude is given by the sum of a series of ladder diagrams
which can be evaluated exactly; this gives us an exact relationship between $g$ and
$g_0$
\begin{eqnarray}
g &=& g_0 \left( 1 + \frac{g_0 \mu^{\epsilon}}{S_d} \int \frac{d^d k}{(2 \pi)^d}
\frac{1}{k^2 + (k+2p)^2} \right)^{-1} \nonumber \\
&\equiv& g_0 \left( 1+ \frac{g_0 A_d}{\epsilon} \right)^{-1}
\label{gg0}
\end{eqnarray}
with
\begin{equation}
A_d \equiv \frac{\epsilon \Gamma (d/2) \Gamma (1-d/2)}{4s^{\epsilon} } = \frac{1}{2}
 - \frac{\ln (s)}{2} \epsilon + {\cal O} (\epsilon^2)
\end{equation}
We can now deduce the exact $\beta$-function of $g$,
\begin{equation}
\beta_g = -\epsilon g + A_d g^2,
\end{equation}
which has a fixed point at $g = g^{\ast} = \epsilon / A_d$.

The effective potential $\Gamma (\psi)$~\cite{domb} (this is the
generating functional of the one particle irreducible vertices) can be easily
determined from the standard Bogoluibov theory of the ground state energy of a dilute
Bose gas:
\begin{equation}
\Gamma (\psi ) = r |\psi|^2 + \frac{u}{2} |\psi|^4 + \frac{1}{2}
\int \frac{d^d k}{(2 \pi )^d} \left[ \left( (k^2 + r + 2 u |\psi |^2 )^2
- u^2 |\psi |^4 \right)^{1/2} - (k^2 + r + 2 u |\psi|^2 ) \right]
\label{bogol} 
\end{equation}
in the one-loop approximation.
To express this result in a universal form, we have to
evaluate the above integral, express $u$ in terms of the renormalized
coupling $g$ using (\ref{ug0},\ref{gg0})
and, finally, set $g$ at its fixed point value $g = g^{\ast}$. 
First, we write (\ref{bogol}) in the form
\begin{eqnarray}
&& \frac{2 \epsilon}{S_d} \Gamma (\psi ) = r |\bar{\psi}|^2 +
\frac{\mu^{\epsilon} g |\bar{\psi}|^4 }{4\epsilon} \left\{ 1 +
g \left( \frac{g|\bar{\psi}|^2}{2\epsilon\mu^{2-\epsilon}}\right)^{-\epsilon/2}
\int_0^{\infty} k^{1-\epsilon} dk \left[ \left( (k^2 + 2\epsilon r \mu^{-\epsilon}
/(g |\bar{\psi}|^2) + 2 )^2 - 1 \right)^{1/2} \right.\right. \nonumber\\
&&~~~~~~~~~~~~~~~~~~~~~~~~~~~~~~~~~~~~~~~~~~~~ - (k^2 +
2 \epsilon r\mu^{-\epsilon} /(g |\bar{\psi}|^2) + 2)\biggr] + \frac{A_d g}{ \epsilon}
\Biggr\},
\end{eqnarray}
where $\bar{\psi} = (2\epsilon/S_d )^{1/2} \psi$, and we expressed $g_0$ only upto
second order in $g$. The above integral can be evaluated in powers of $\epsilon$; as
expected, the poles in $\epsilon$ within the curly brackets cancel. Then, set $g =
g^{\ast} = 
\epsilon/A_d$ and expand the whole expression to order $\epsilon$. The $\mu$ and the
$s$ dependence  disappears, and the resulting expression is completely universal.
We can write it in the scaling form
\begin{equation}
\Gamma (\psi ) = |r|^{1 + d/2} \Phi \left( \frac{2 \epsilon |\psi|^2}{
S_d |r|^{d/2}} \mbox{sgn}(r) \right)
\label{zeroscale}
\end{equation}
where $\Phi$ is a universal scaling function, and all exponents are written in
their expected exact form. Notice that there are no arbitrary scale factors in
(\ref{zeroscale})---this is the no scale-factor universality of Ref~\cite{sss}.
The scaling function $\Phi (y)$ is determined by the above calculation to
order $\epsilon$:
\begin{eqnarray}
&&\Phi (y) = \frac{S_d}{2 \epsilon} \left[
y + \frac{y^2}{2} + \frac{\epsilon}{4} \left(
y^2 \log \left( \frac{\mbox{sgn}(y)(1 + 2y) + (1 + 4 y + 3 y^2)^{1/2}}{2} \right)
\right.\right.
\nonumber \\
&&~~~~~~~~~~~~~~~~~~~~~~~~~~~~~~~~~
\left.\left. - \mbox{sgn}(y)(1+2y) (1 + 4y + 3 y^2)^{1/2} + 1 + 4 y +
\frac{7y^2}{2}
\right) + {\cal O}(\epsilon^2) \right]
\end{eqnarray}

The condensate $\psi_0 = \langle \psi \rangle$ is determined by the condition
$\partial \Gamma/ \partial \psi |_{\psi=\psi_0} = 0$, while the total density of
particles, $n$, is given by $n = \partial \Gamma / \partial r$. Using (\ref{zeroscale})
this gives us
\begin{equation}
n = \mbox{sgn}(r) |r|^{d/2} \left(1 + \frac{d}{2} \right) \Phi (y_0)~~~~~~~~
\mbox{where $y_0 \Phi^{\prime} (y_0 )  = 0$}
\end{equation}
This gives a result for the density in the form (\ref{ncalc}) with
\begin{equation}
{\cal C}_d = S_d \left( \frac{1}{2 \epsilon} - \frac{1 - \log 2}{4} 
+ {\cal O}(\epsilon ) \right).
\label{rescalcd}
\end{equation} 
Finally, we recall that in $d=1$, the exact value of ${\cal C}_d$ is
known~\cite{sss}: ${\cal C}_1 = S_1$.

\section{Computations for Quantized Ferromagnet (B) to Canted (D) Transition}
\label {apprg2}
At the critical point, the propagator of (\ref{actionbtodn}) is (dropping the
$0$ subscripts on the couplings)
\begin{eqnarray}
G(x,\tau) &=& \int \frac{d^d k}{(2 \pi)^d} \frac{d \omega}{2 \pi}~
\frac{(-i M_0 \omega + k^2 )e^{i (\vec{k} \cdot \vec{x} - \omega
\tau)}}{-\omega^2 M_0^2 \lambda - i M_0 \omega k^2 (1 +
\lambda ) + k^4 ( 1 - b)}
\nonumber \\ &=& 
\frac{M_0^{(d-2)/2}}{(4 \pi |\tau|)^{d/2}} \left(A_{+} \frac{e^{-M_0 x^2 /4
\lambda_+ \tau}}{
\lambda_+^{d/2}} \theta (\tau) - A_{-} 
\frac{ e^{-M_0 x^2 /4 \lambda_- |\tau|}}{
\lambda_-^{d/2}} \theta (-\tau) \right)
\end{eqnarray}
where
\begin{equation}
\lambda_{\pm} = \frac{ -((1+\lambda)^2 -4\lambda(1-b))^{1/2} \pm (1+\lambda)}{2
\lambda},
\end{equation}
and
\begin{equation}
A_{\pm} = \frac{\lambda_{\pm} \mp 1}{\lambda ( \lambda_{+} + 
\lambda_{-})}.
\end{equation}
We have assumed above, and in the remainder of this Appendix that $\lambda < 0$.

The order $u^2$ contribution to the self energy is therefore
\begin{equation}
\Sigma (k, \omega) = 2 u^2 \int d^d x d \tau G^2 (x, \tau) G(-x, -\tau)
e^{-i (\vec{k} \cdot \vec{x} - \omega \tau)}
\end{equation}
After a lengthy, but straightforward, evaluations of the above integral
we find, to leading order in $\epsilon = 2-d$:
\begin{equation}
\Sigma (k, \omega ) - \Sigma (0,0) = 
- \frac{u^2 }{8 \pi^2 M_0^2 \lambda^2 \epsilon} ( k^2 R_1 (b,\lambda) 
- i M_0 \omega R_2 (b, \lambda) ),
\label{apprg2res1}
\end{equation}
and, as in (\ref{calcz4}), the renormalized four-point vertex
\begin{equation}
u - \frac{u^2}{4 \pi M_0 |\lambda| \epsilon} R_3 (b, \lambda),
\label{apprg2res1a}
\end{equation}
where
\begin{eqnarray}
R_1(b, \lambda) &=& \frac{ b \lambda^2 ((1+ \lambda)^2
+ 3(1-b)(1-2\lambda))}{((1+\lambda)^2 - 4 \lambda
(1-b))(2 \lambda^2 + 9 b \lambda - 5 \lambda + 2)^2}
\nonumber \\
R_2(b, \lambda) &=& \frac{b \lambda^2 (\lambda + 3 b - 2)}{(1-b)((1+\lambda)^2 - 4
\lambda (1-b))(2 \lambda^2 + 9 b \lambda - 5 \lambda + 2)} \nonumber \\
R_3 (b, \lambda) &=& \frac{(1-\lambda)^3 - b(1 + 10 \lambda + 5
\lambda^2 ) + 12 b^2 \lambda}{(1-b)((1+\lambda)^2 - 4
\lambda (1-b))^{3/2}}.
\label{apprg2res2}
\end{eqnarray}
To determine the renormalization of $|\psi|^2$ insertions we need the vertex
between a $|\psi|^2$ operator and a $\psi$ and a $\psi^{\ast}$; this is 
\begin{equation}
1 - \frac{u}{2 \pi M_0 |\lambda| \epsilon} R_4 (b, \lambda),
\label{apprg2res1b}
\end{equation}
where
\begin{equation}
R_4 (b, \lambda) = \frac{-4 b \lambda}{((1+\lambda)^2 - 4
\lambda (1-b))^{3/2}} .
\label{apprg2res3}
\end{equation} 
These results lead immediately to the renormalization constants and 
$\beta$ functions in Section~\ref{secrgBtoD}b for $\lambda<0$.



\begin{references}
\bibitem{tops} We could also have studied an analogous ``quantum top
model'' to
allow for the possibility of
phases in which the spins are non-collinear but have no net ferromagnetic
moment---the analog
of the N\'{e}el phase in that model would have 3 spin-wave modes, all
with
$\omega \sim k$.
The phases with a ferromagnetic moment in the quantum top model are
quite
similar to those being
discussed here in the  quantum rotor model, and we will therefore not
enter
into this complication in
this paper.
\bibitem{limits} The limits of momentum and frequency going to zero do not commute
in the computation of the effective coupling of $\phi$ to $H_z$; it is necessary to
generalize to a momentum- and frequency-dependent field $H_z (q, \Omega)$ and take
the limit $\Omega \rightarrow 0$ before taking $q \rightarrow 0$.
\bibitem{foot} Because systems with $q>0$ do not have ${\cal P}$
symmetry, terms which couple
the `spin' field $\mbox{Tr} ( \vec{\sigma} g)$ (in the notation of
Ref~\protect\cite{affleckhaldane}) with the magnetization density
$\vec{J}_L + \vec{J}_R$
are also permitted. It may be necessary to include this term in the most
general case.
\end{references}
\end{document}